\documentclass[floatfix,twocolumn,showpacs,preprintnumbers,amsmath,amssymb,pra,superscriptaddress,longbibliography]{revtex4-1}
\usepackage{color}
\usepackage[usenames,dvipsnames,svgnames,table]{xcolor}
\usepackage[colorlinks=true,linkcolor=blue,urlcolor=blue,citecolor=blue]{hyperref}

\usepackage{mathtools}
\usepackage{graphicx}
\usepackage{dcolumn}
\usepackage{array}
\usepackage{lipsum}
\usepackage{bm}
\usepackage{subfigure}
\usepackage{amssymb}
\usepackage{multirow}
\usepackage{tabularx}
\usepackage{amsmath}
\usepackage{braket}

\renewcommand{\vec}[1]{\mathbf{#1}}

\begin{document}

\title{ Linear-response time-dependent density functional theory approach to warm dense matter with adiabatic exchange--correlation kernels}

\author{Zhandos A. Moldabekov}
\email{z.moldabekov@hzdr.de}
\affiliation{Center for Advanced Systems Understanding (CASUS), D-02826 G\"orlitz, Germany}
\affiliation{Helmholtz-Zentrum Dresden-Rossendorf (HZDR), D-01328 Dresden, Germany}

\author{Michele Pavanello}
\affiliation{Department of Chemistry, Rutgers University, 73 Warren St., Newark, NJ 07102, USA}
\affiliation{Department of Physics, Rutgers University, 101 Warren St., Newark, NJ 07102, USA}

\author{Maximilian~P.~B\"ohme}
\affiliation{Center for Advanced Systems Understanding (CASUS), D-02826 G\"orlitz, Germany}
\affiliation{Helmholtz-Zentrum Dresden-Rossendorf (HZDR), D-01328 Dresden, Germany}

\affiliation{Technische  Universit\"at  Dresden,  D-01062  Dresden,  Germany}

\author{Jan Vorberger}
\affiliation{Helmholtz-Zentrum Dresden-Rossendorf (HZDR), D-01328 Dresden, Germany}

\author{Tobias Dornheim}
\affiliation{Center for Advanced Systems Understanding (CASUS), D-02826 G\"orlitz, Germany}
\affiliation{Helmholtz-Zentrum Dresden-Rossendorf (HZDR), D-01328 Dresden, Germany}

\begin{abstract}
We present a new methodology for the linear-response time-dependent density functional theory (LR-TDDFT) calculation of the dynamic density response function of warm dense matter in  an adiabatic approximation that can be used with any available exchange-correlation (XC) functional across Jacob's Ladder and across temperature regimes. 
The main novelty of the presented approach is that it can go beyond the adiabatic local density approximation (ALDA) and generalized LDA (AGGA)  while preserving the self-consistence between the Kohn-Sham (KS) response function and adiabatic XC kernel for extended systems.  The key ingredient for the presented method is the combination of the adiabatic XC kernel from the direct perturbation approach with the  macroscopic dynamic KS response from the standard LR-TDDFT method using KS orbitals.  We demonstrate the application of the method for the example of warm dense hydrogen, for which we perform a detailed analysis of the KS density response function, the RPA result, the total density response function and of the adiabatic XC kernel. 
The analysis is performed using LDA, GGA, and meta-GGA level approximations for the XC effects. The presented method is directly applicable to disordered systems such as liquid metals, warm dense matter, and dense plasmas. 

\end{abstract}

\maketitle

\section{Introduction} 

Warm dense matter (WDM) is relevant for a plethora of applications such as inertial confinement fusion (ICF) \cite{Zylstra2022, Betti2016}, laboratory astrophysics \cite{Kritcher2020, Kraus2017}, and material science at extreme conditions \cite{Lazicki2021, Kraus2016, Schuster_diamond_2022}. WDM is generated in experiments e.g.~by laser heating and/or shock compression at facilities such as the National Ignition Facility (NIF) \cite{Betti2016,Zylstra2022, hu_ICF}, the European X-ray Free-Electron Laser (XFEL) \cite{Tschentscher_2017}, and the Linac coherent light source (LCLS) at SLAC \cite{He2022}. In nature, WDM exists in the interior of white dwarfs \cite{CBFS00, saumon1} and giant planets \cite{Schuster_PRB, SEJ14}, and in the outer layer of neutron stars \cite{Daligault_2009}.   
Among various atomic composition possibilities, warm dense hydrogen is of particular importance. Indeed, hydrogen is the most abundant element in the universe and its isotopes are used as thermonuclear fuel in the ICF. Recently, a breakthrough achievement on the path to ICF has been reported: a net positive energy outcome from fusion reactions compared to the energy used to start the fusion process has been detected. This makes the understanding of warm dense hydrogen properties and the development of reliable methods for WDM~\cite{wdm_book,new_POP,POP_review}, in general, highly relevant. 

At the extreme temperatures and densities realized in WDM experiments and within the aforementioned astrophysical objects, electrons are partially degenerate and strongly correlated. With regard to temperature and density,  WDM occupies an intermediate region between solid state and dense plasmas. This is also reflected in the theoretical methods used for WDM research~\cite{POP_review}, some of which we discuss in this work. Being a relatively new field compared to solid state physics and plasma physics, theoretical methods to describe WDM are still in the stage of active development. Among existing approaches, \textit{ab initio} methods such as quantum Monte-Carlo (QMC)~\cite{review,dornheim_sign_problem,Driver_PRL_2012,wdm_book} and thermal density functional theory (DFT)~\cite{POP_review,Pribram_Thermal_Context_2014,karasiev_importance,kushal} are particularly important for a reliable description of  experimental results and for providing guide for new experimental developments.

One of the main theories for the description of many-particle systems is the linear response theory (LRT). Although basic principles of the LRT are general, the LRT is formulated in somewhat different ways and with emphasis on different physical features in solid state physics, in WDM studies, and in plasma physics. On the one hand, WDM is usually disordered and does not feature symmetries and a well defined crystal structure like solids. On the other hand, strong electron-ion and electron-electron correlations do not allow neglecting microscopic density inhomogeneities as it is often the case for plasmas. Additionally, for WDM it is important to have access to the density response properties at wave numbers comparable or larger than the Fermi wave number since such wave numbers are probed by the x-ray Thomson scattering (XRTS) technique~\cite{siegfried_review,sheffield2010plasma}, which is one of the main WDM diagnostics methods~\cite{Gregori_PRE_2003,kraus_xrts,Dornheim_T_2022,Dornheim_T_follow_up}. It is crucial to have a clear understanding of these aspects when tools developed for solids or plasmas are used for WDM studies. 

An important example of such a tool is the linear-response time-dependent density functional theory (LR-TDDFT) approach. LR-TDDFT allows one to compute  various electronic optical and transport properties~\cite{marques2012fundamentals}. The key quantity for the LR-TDDFT based description is the density response function.
Indeed, an accurate computation of the density response function is one of the main problems of the WDM theory~\cite{POP_review,Hamann_PRB_2020}.

A quantity that is closely connected to  the density response function is the exchange-correlation (XC) kernel, which can be defined as the functional derivative of the XC potential with respect to density \cite{Gross_PRL1985}. The XC kernel is one of the key ingredients for the computation of the density response function using the LR-TDDFT.
Recently, Moldabekov {\textit et al.}~\cite{Moldabekov_dft_kernel} have presented an approach for the calculation of the static (adiabatic) XC kernel within KS-DFT for any available XC functional. This method is formulated for disordered systems that are homogeneous on average, e.g. WDM, dense plasmas, and liquid metals.
This method is based on the direct perturbation of the system by a static harmonic potential~\cite{moroni,moroni2,bowen2,Dornheim_PRL_2020,Moldabekov_JCTC_2022,POP_review}. The difference in the density between perturbed and unperturbed cases allows one to extract the static density response function. This static density response function is combined with a reference non-interacting density response function to compute the static XC kernel. 
The main utility of the static XC kernel within LR-TDDFT is the computation of the density response function. However, it had hitherto not been clear how to use the static XC kernel from the direct perturbation approach in the LR-TDDFT framework.
In this work, we formulate a consistent LR-TDDFT approach for disordered systems such as WDM that is based on the adiabatic (static)  XC kernel obtained from the direct perturbation approach.

As a the main result, we formulate a clear recipe for the computation of the  linear density response function combining both the LR-TDDFT and direct perturbation method. 
Within this framework, the LR-TDDFT is used to compute the macroscopic  KS-response function taking into account the density inhomogeneity induced by ions  (\textit{local-field effects} \cite{Wiser, Adler}) and the direct perturbation method provides access to the static XC kernel for any XC functional. These two ingredients are combined to compute the macroscopic linear density response function of WDM. 
We apply this combined workflow to investigate density response properties and the XC kernel of warm dense hydrogen. 
We present a detailed numerical investigation of the density response function, the Kohn-Sham (KS) response function and its screened version in the random phase approximation (RPA), and of the exchange-correlation (XC) kernel. To get insights into the role of XC effects, we use  XC functionals on the level of the local density approximation (LDA), generalized gradient approximation (GGA) and meta-GGA. Although we mainly focus on WDM applications, the presented workflow is applicable for any disordered system that is homogeneous on average.

The paper is organized as follows: we begin with providing the theoretical background of LR-TDDFT, of LRT as it is used for homogeneous systems (such as plasmas), and of the the direct perturbation approach in Sec.~\ref{s:theory}; we conclude Sec.~\ref{s:theory} by formulating an adiabatic approximation for the density response function based on the static XC kernel from the direct perturbation approach.
The parameters and simulation details are given in Sec.~\ref{s:smd}, and we illustrate the application of the developed method for the example of warm dense hydrogen in Sec.~\ref{s:all_results}. The paper is concluded by a summary of the main findings and a brief outlook on future applications.


\section{Theoretical background}\label{s:theory}


Different formulations and approaches are being used for the calculation and description of the  linear density response function of electrons in WDM.
First of all, there is the standard LR-TDDFT approach. It uses the Kohn-Sham (KS) density response function, which is based on KS orbitals, and the exchange-correlation kernel \cite{book_Ullrich}. In this case, the impact of ions on electrons is taken into account in both the KS response function and exchange-correlation kernel through the density inhomogeneity induced by the ions (local-field effects).
Another often used approach, originally developed for dense plasma physics, is based on the ideal Lindhard response function and the so-called local field correction~\cite{kugler1}, which incorporates everything else that is missing in the Lindhard function (not to be confused with local-field effects in the LR-TDDFT) \cite{Fortmann_PRE_2010, Hamann_PRB_2020, Reinholz_PRE, Hamann_CPP_2020}.   This is not only a standard plasma physics approach, but also commonly used in the quantum theory of the homogeneous electron liquid \cite{quantum_theory, kugler1}. From a computational point of view, an alternative to the LR-TDDFT is to use a direct perturbation by an external field to measure the density response function connecting the perturbing field with the density change induced by it \cite{  direct_app_1984, direct_app_1985, Sakko}. These different formulations and methods for a dynamic density response function have been actively used in the WDM studies \cite{Moldabekov_JCTC_2022, hybrid_results, Moldabekov_non_empirical_hybrid, dynamic2, ramakrishna2022electrical}. However, in many aspects, the connection between them had not been clearly formulated and understood.

The main goal of this work is to formulate a consistent LR-TDDFT approach for WDM that is based on an adiabatic (static) XC kernel from the direct perturbation approach introduced in Ref~\cite{Moldabekov_dft_kernel}. 
We will first discuss the dynamic dielectric function from LR-TDDFT, followed by an introduction to LRT as it is used in the physics of dense plasmas, and the direct perturbation approach to the static XC kernel. This allows us to present a clear connection between these schemes and a recipe for LR-TDDFT calculations of the macroscopic density response of WDM with an adiabatic XC kernel on any rung on Jacob's ladder.


\subsection{Dynamic dielectric function from LR-TDDFT}\label{s:lr-tddft}
We start with a brief summary of  what is needed from LR-TDDFT for further discussion of the connection between the latter and the direct perturbation approach.

The microscopic dynamic dielectric function, for  the momentum $\vec k$  and energy $\omega$, is expressed in terms of the microscopic density response function as~\cite{book_Ullrich}
\begin{equation}\label{eq:d_f}
      \varepsilon^{-1}_{\scriptscriptstyle \vec G,\vec G^{\prime}}(\vec k,\omega)=\delta_{\scriptscriptstyle \vec G,\vec G^{\prime}}+\frac{4\pi}{\left|\vec k+\vec G\right|^{2}}  \chi_{\scriptscriptstyle \vec G,\vec G^{\prime}} (\vec k,\omega),
\end{equation}
where  $\vec k$ is a wave vector restricted to the first Brillouin zone, and  $\vec G$ and $\vec G^{\prime}$ are reciprocal lattice vectors.

The Fourier coefficients $  \chi_{\scriptscriptstyle \vec G,\vec G^{\prime}}(\vec k,\omega)$ of the density response matrix are computed using a Dyson’s type equation  \cite{Byun_2020, Adragna_PRB, martin_reining_ceperley_2016}:
\begin{equation}\label{eq:Dyson}
\begin{split}
\chi_{\scriptscriptstyle \mathbf G \mathbf G^{\prime}}(\mathbf k, \omega)
&= \chi^0_{\scriptscriptstyle \mathbf G \mathbf G^{\prime}}(\mathbf k, \omega)+ \displaystyle\smashoperator{\sum_{\scriptscriptstyle \mathbf G_1 \mathbf G_2}} \chi^0_{\scriptscriptstyle \mathbf G \mathbf G_1}(\mathbf k, \omega) \big[ v_{\scriptscriptstyle \mathbf G1}(\vec k)\delta_{\scriptscriptstyle \mathbf G_1 \mathbf G_2} \\
&+ K^{\rm xc}_{\scriptscriptstyle \mathbf G_1 \mathbf G_2}(\mathbf k, \omega) \big]\chi_{\scriptscriptstyle \mathbf G_2 \mathbf G^{\prime}}(\mathbf k, \omega),
\end{split}
\end{equation}
where   $\chi^{~0}_{\scriptscriptstyle \vec G,\vec G^{\prime}}(\vec k,\omega)$ is a non-interacting density response function, $v_{\scriptscriptstyle \mathbf G1}(\vec k)={4\pi}/{|\mathbf k+\mathbf G_1|^2}$ is the Coulomb potential, and $ K^{\rm xc}_{\scriptscriptstyle \vec G_1,\vec G_2}(\vec k, \omega)$ is the the exchange-correlation kernel. The non-interacting density response function  $\chi^{~0}_{\scriptscriptstyle \vec G,\vec G^{\prime}}(\vec k,\omega)$ is computed using the KS-orbitals and corresponding energy eigenvalues from a KS-DFT simulation of an equilibrium state \cite{Hybertsen}.

There are numerous approximations for the exchange-correlation kernel in Eq.~(\ref{eq:Dyson}), which we do not discuss in this work (see, e.g., Refs.~\cite{Onida_RevModPhys, Byun_2020} for more information).  We only note that the exchange-correlation kernel is defined as the functional derivative of the XC potential with respect to density\cite{book_Ullrich}.

A commonly used approximation for the XC kernel that is considered in this work  is the adiabatic (static) approximation $ K^{\rm xc}_{\scriptscriptstyle \vec G_1,\vec G_2}(\vec k)=K^{\rm xc}_{\scriptscriptstyle \vec G_1,\vec G_2}(\vec k, \omega=0)$. For extended systems with sampling of the  Brillouin-zone, currently there is no possibility to compute $K^{\rm xc}_{\scriptscriptstyle \vec G_1,\vec G_2}(\vec k)$  using the functional derivative of the XC potential beyond the adiabatic LDA (ALDA) and GGA (AGGA)~\cite{Byun_2020}. 
This  is  meant in the sense of the self-consistent calculation where the XC kernel is fully compatible  with the XC functional used for the KS-DFT simulation of the equilibrium state. 


The response of a system to an external monochromatic perturbation with wave vector $\vec q=\vec k +\vec G$ (note that $\vec q$ is not restricted to  the first Brillouin zone like $\vec k$), e.g. in the case of the XRTS diagnostics, is described by the value of the microscopic dynamic dielectric function (\ref{eq:d_f}) with $\vec G=\vec G^{\prime}$ ($\vec G=\vec G^{\prime}=0$, if $\vec q$ is in the first Brillouin zone) \cite{DSF_LR-TDDFT, martin_reining_ceperley_2016}:
\begin{equation}\label{eq:d_f_lf}
    \varepsilon_M(\vec q,\omega)=\frac{1}{1+\frac{4\pi}{\left|\vec q\right|^{2}}  \chi_{\scriptscriptstyle \vec G\vec G} (\vec k,\omega)}.
\end{equation}

Further, following the LR-TDDFT terminology, we call the physical properties describing the response of a system to an external macroscopic perturbation  as macroscopic; e.g.  $\varepsilon_M(\vec q,\omega)$. 


The random phase approximation (RPA)  follows from  setting  $ K^{\rm xc}_{\scriptscriptstyle \vec G_1,\vec G_2}(\vec k, \omega)=0$ in Eq.~(\ref{eq:Dyson}): 

\begin{equation}\label{eq:Dyson_rpa}
\begin{split}
\chi^{\rm RPA}_{\scriptscriptstyle \mathbf G \mathbf G^{\prime}}(\mathbf k, \omega)
&= \chi^0_{\scriptscriptstyle \mathbf G \mathbf G^{\prime}}(\mathbf k, \omega)\\
&+ \sum_{\scriptscriptstyle \mathbf G_1 } \chi^0_{\scriptscriptstyle \mathbf G \mathbf G_1}(\mathbf k, \omega) \frac{4\pi}{|\mathbf k+\mathbf G_1|^2}
\chi^{\rm RPA}_{\scriptscriptstyle \mathbf G_1 \mathbf G^{\prime}}(\mathbf k, \omega),
\end{split}
\end{equation}

Accordingly, the macroscopic dielectric function in the RPA reads:
\begin{equation}\label{eq:d_f_RPA_lf}
    \varepsilon^{\rm RPA}_M(\vec q,\omega)=\frac{1}{1+\frac{4\pi}{\left|\vec q\right|^{2}}  \chi^{\rm RPA}_{\scriptscriptstyle \vec G\vec G} (\vec k,\omega)}.
\end{equation}




Next, we discuss the LRT formulation as it is used for homogeneous systems such as WDM and dense plasmas.

\subsection{Linear density response theory in dense plasma physics}\label{s:plasma}

In the LR-TDDFT, ions are considered to be an external field that affects electronic properties. 
In contrast, the linear response theory of plasmas is generally formulated taking into account both the response of ions and electrons to an external perturbation \cite{Ichimaru_PRA}.
To be on the same level of description with the LR-TDDFT, we consider the response of electrons with fixed ions. 
We note that this is justified in situations where the perturbation time is much shorter than the reaction time of ions since the latter are much more inert compared to electrons \cite{S_Vinko}.

The dielectric function of electrons in homogeneous systems is expressed using the density response function of electrons $\chi(\vec q,\omega)$ in the following form \cite{quantum_theory}:
\begin{equation}\label{eq:df_plasma}
   \varepsilon^{-1}(\vec q,\omega)=1+v(q)\chi(\vec q,\omega),
\end{equation}
where $v(q)=4\pi/q^2$.

It is a common practice in the quantum theory of the electron liquid and in dense plasma physics to 
express the density response function in terms of  the ideal response function and  the XC kernel \cite{quantum_theory, plasma2, Ichimaru_PRA} : 
\begin{eqnarray}\label{eq:kernel}\label{eq:chi_p}
 \chi(\mathbf{q},\omega) = \frac{\chi_0(\mathbf{q},\omega)}{1 -\left[v(q)+K_{\rm xc}(\mathbf{q},\omega)\right]\chi_0(\mathbf{q},\omega)},
\end{eqnarray}
where $\chi_0(\vec q, \omega)$ is a reference non-interacting density response function   and the XC kernel
often defined using so called local field correction (LFC):
\begin{eqnarray}\label{eq:LFC}
 K_\textnormal{xc}(\mathbf{q},\omega)=-v(q) G(\mathbf{q},\omega)\ .
\end{eqnarray}

We note that in the case of the uniform electron gas (UEG), the XC kernel in Eq.~(\ref{eq:chi_p}) is the same as the XC kernel of the LR-TDDFT.
However, in general, for disordered systems that can be considered as homogeneous after averaging over different ionic configurations---such as WDM, dense plasmas, and liquid metals---a clear connection between the XC kernel in Eq.~(\ref{eq:chi_p}) and the microscopic XC kernel in the LR-TDDFT (e.g., in Eq.~(\ref{eq:Dyson})) had not been provided.

The  dielectric function in the RPA follows from setting $K_{\rm xc}(\mathbf{q},\omega)=0$ in Eq.~(\ref{eq:chi_p}) and substituting the resulting $ \chi(\mathbf{q},\omega)$ into Eq.~(\ref{eq:df_plasma}):
\begin{equation}\label{eq:df_plasma_rpa}
   \varepsilon_{\rm RPA}(\vec q,\omega)=1-v(q)\chi_0(\vec q,\omega).
\end{equation}

In plasma physics applications, it is common to use the  Lindhard function (temperature-dependent version) as $\chi_0(\vec q, \omega)$ and approximate $G(\vec q,\omega)$ according to free electron gas model \cite{Ichimaru_PRA, Fortmann_PRE_2010}.


For WDM applications, the correction due to electron-ion interactions  to $\varepsilon(\vec q,\omega)$ is often included explicitly in the form of the electron-ion collision  frequency in the so-called generalized Mermin model \cite{Sperling_PRL_2015, Fortmann_PRE_2010, Witte_PRL_2017, plagemann2012dynamic, siegfried_review}. The relevant point is that this model is also based on the temperature-dependent Lindhard function and  $G(\vec q,\omega)$.
The Lindhard function for $\chi_0(\vec q, \omega)$ can be used when the composition of the system can be divided into well defined species like free electrons, ions, and neutral atoms.
This is often the case for plasmas, where a chemical model like Saha equation is used for the computation of the free electron density \cite{kremp2005quantum}.
However, in WDM it is usually not possible to clearly segregate free and bound electrons and, thus, any concept based on the Lindhard function (i.e., free electron gas) requires introduction of the effective free electron density (or an effective ionization degree). This is commonly used as a fit parameter to describe experimental XRTS data  \cite{Sperling_PRL_2015, Witte_PRL_2017, plagemann2012dynamic, siegfried_review}. We further note that the effective free electron density is also used to compute $G(\mathbf{q},\omega)$ on the basis of some free electron gas model.  In contrast, the LR-TDDFT approach described in Sec. \ref{s:lr-tddft} does not require  the effective free electron density. Moreover, electron-ion interaction effects are included in the LR-TDDFT implicitly through microscopic density inhomogeneity and KS-orbitals. Therefore, it is highly beneficial to reformulate Eq.~(\ref{eq:chi_p}) without using the  Lindhard function and $G(\mathbf{q},\omega)$ (equivalently $ K_\textnormal{xc}(\mathbf{q},\omega)$) of the free electron gas, but using some $\chi_0(\vec q, \omega)$ and $ K_\textnormal{xc}(\mathbf{q},\omega)$ from the LR-TDDFT. This we do after introducing the method to compute the static  XC kernel and KS response function using a harmonic perturbation in the KS-DFT calculations.   


\subsection{Static XC kernel from the direct perturbation approach}\label{s:direct_per}

The direct perturbation approach is based on the KS-DFT simulation of a  system perturbed by an external harmonic potential \cite{Hybertsen, direct_app_1985, direct_app_1984}.
The corresponding electronic Hamiltonian reads~\cite{moroni,moroni2,Dornheim_PRL_2020,POP_review}
\begin{eqnarray}\label{eq:Hamiltonian_modified}
 \hat{H}_{\mathbf{q},A} = \hat{H}_e + 2 A \sum_{j=1}^N \textnormal{cos}\left( \mathbf{q}\cdot\hat{\mathbf{r}}_j \right)\ ,
\end{eqnarray}
where  $\hat{H}_e$ is the Hamiltonian of the unperturbed system. The static external potential is defined by its wave vector $\mathbf{q}$ and amplitude $A$. 
The discussion of the general case of a time dependent external potential can be found in Ref.~\cite{POP_review}.

The simulation of both the unperturbed ($A=0$) and perturbed ($A\neq 0$) systems allows one to find  a change in the single-electron density:
\begin{eqnarray}\label{eq:delta_n}
 \Delta n_e(\mathbf{r})_{\mathbf{q}, A} = \braket{n_e(\mathbf{r})}_{\mathbf{q}, A} - \braket{n_e(\mathbf{r})}_{A=0}\ .
\end{eqnarray}

In WDM experiments, the XRTS measurements are performed on macroscopic objects. With regard to a  macroscopic disordered system, what the x-ray probe ``sees'' statistically is a large ensemble of atomic position configurations.   Therefore, in practice, we perturb a system along a given direction and compute the average $\braket{...}$ along all other directions. Depending on the system  temperature and density, one might need further averaging over different atomic configurations. We note that it does not mean that we neglect the local field effects (see Sec. \ref{s:connection}).
For such disordered systems, the contributions from the density Fourier components to $\Delta n_e(\mathbf{r})_{\mathbf{q}, A}$ at $\vec q+\vec G$, $\vec q+2\vec G$ \textit{etc.} are negligible \cite{Moldabekov_dft_kernel}. Physically, it becomes obvious if we recall that for disordered systems (like molten metals) $\vec G$ does not describe the Fourier transform of the Bravais lattice, but appears due to periodic boundary conditions (with $\left|\vec G \right|=2\pi/L$) introducing a certain finite size effect, which we discuss in Sec. \ref{s:smd} and Sec. \ref{s:all_results}.  Furthermore, if the perturbation amplitude $A$ is small enough, one can also neglect the excitation of higher harmonics due to  the non-linear response terms at $2\vec q$, $3\vec q$ \textit{etc.} \cite{Bohme_PRL_2022, Dornheim_PRR_2021, Moldabekov_JCTC_2022}. Therefore,  the induced density change is  described by the  linear density response function $\chi(\mathbf{q}) = \chi(\mathbf{q},\omega=0)$ in the static limit according to the relation~\cite{Dornheim_PRR_2021}
\begin{eqnarray}\label{eq:delta_n_LRT}
   \Delta n_e(\mathbf{r})_{\mathbf{q},A}  = 2 A \textnormal{cos}\left(\mathbf{q}\cdot\mathbf{r}\right)\chi(\mathbf{q})\  .
\end{eqnarray}

Similarly, a static KS response function describes the response of the electron density to a change of the KS potential:
\begin{eqnarray}\label{eq:delta_KS}
   \Delta n_e(\mathbf{r})_{\mathbf{q},A}  = \chi_{\rm KS}(\mathbf{q}) \Delta v_{\rm KS}(\mathbf{r})_{\mathbf{q},A}\  ,
\end{eqnarray}
where $\Delta v_{\rm KS}(\mathbf{r})_{\mathbf{q},A}$ is the change of the KS potential due to an external perturbation:
\begin{eqnarray}\label{eq:delta_v}
 \Delta v_{\rm KS}(\mathbf{r})_{\mathbf{q}, A} = \braket{v_{\rm KS}(\mathbf{r})}_{\mathbf{q}, A} - \braket{v_{\rm KS}(\mathbf{r})}_{A=0}\ .
\end{eqnarray}

Therefore, the direct perturbation approach provides access to both macroscopic density response function and KS response function of a homogeneous system. 
We reiterate that this method is valid for any disordered system that becomes homogeneous on average; e.g. due to averaging over an ensemble of atomic configurations (snapshots).


Now, we can set  $\chi_0(\vec q, \omega)=\chi_{\rm KS}(\vec q, \omega)$  and invert Eq.~(\ref{eq:chi_p}) to find a static XC kernel for a homogeneous system \cite{quantum_theory}:

\begin{equation}\label{eq:invert}
\begin{split}
     K_\textnormal{xc}(\mathbf{q}) &= -\left\{
 v(q) + \left( \frac{1}{\chi(\mathbf{q})} - \frac{1}{\chi_{\rm KS}(\mathbf{q})} \right)
 \right\}\ \\
 &= \frac{1}{\chi_\textnormal{RPA}(\mathbf{q})} - \frac{1}{\chi(\mathbf{q})}\ ,
\end{split}
\end{equation}
where $\chi_\textnormal{RPA}(\mathbf{q})$ is the screened version of  $\chi_{\rm KS}(\mathbf{q})$: 
\begin{eqnarray}\label{eq:kernel}\label{eq:chi_p_rpa}
 \chi_{\rm RPA}(\mathbf{q},\omega) = \frac{\chi_{\rm KS}(\mathbf{q},\omega)}{1 -v(q)\chi_{\rm KS}(\mathbf{q},\omega)}.
\end{eqnarray}


\subsection{Connection between LR-TDDFT and direct perturbation approach}\label{s:connection}


To access the macroscopic density response function $\chi(\mathbf{q})$  (defined according to Eq.~(\ref{eq:delta_n_LRT})) without applying an external harmonic perturbation, we use the macroscopic dielectric function Eq.~(\ref{eq:d_f_lf}) in Eq.~(\ref{eq:df_plasma}) to find the LR-TDDFT result for $\chi(\mathbf{q})$:

\begin{equation}\label{eq:chi_lf}
    \chi(\vec q,\omega)=\frac{1}{v(q)}\left({\varepsilon^{-1}_M}(\vec q, \omega)-1\right).
\end{equation}

Using Eq.~(\ref{eq:chi_lf}), now we can compute the  macroscopic static density response function $\chi(\vec q)=\chi(\vec q,\omega=0)$ using the LR-TDDFT and compare it with the $\chi(\vec q)$ from the direct perturbation approach.  In Sec. \ref{s:tddft_vs_dp}, we show numerical results confirming that the $\chi(\vec q)$ from the direct perturbation approach Eq.~(\ref{eq:delta_n_LRT}) is equivalent to the LR-TDDFT result computed using Eq.~(\ref{eq:chi_lf}).

To get access to the macroscopic KS response function $\chi_{\rm KS}(\vec q,\omega)$ (defined according to Eq.~(\ref{eq:delta_n_LRT}))  without perturbing by an external harmonic potential, we perform  the LR-TDDFT calculations without XC kernel and use the macroscopic dielectric function in the RPA $ \varepsilon^{\rm RPA}_{ M}(\vec q,\omega)$  from Eq.~(\ref{eq:d_f_RPA_lf}) in Eq.~(\ref{eq:df_plasma_rpa}): 

\begin{equation}\label{eq:macro_ks}
    \chi_{\rm KS}(\vec q,\omega)=\frac{1}{v(q)}\left(1-\varepsilon^{\rm RPA}_M(\vec q, \omega)\right).
\end{equation}

We note that the macroscopic KS response function $\chi_{\rm KS}(\vec q,\omega)$ introduced by Eq.~(\ref{eq:macro_ks}) involves the solution of the Dyson type equation (\ref{eq:Dyson_rpa}). In the LR-TDDFT language, this means that $\chi_{\rm KS}(\vec q,\omega)$ has more information about ions induced density inhomogeneity compared to the non-interacting density response function $\chi^{0}_{GG}(\vec q,\omega)$. To the best of  our  knowledge, the macroscopic KS response function defined by  Eq.~(\ref{eq:macro_ks}) had not been presented in prior works. In Sec. \ref{s:tddft_vs_dp}, we compute $\chi_{\rm KS}(\vec q)$ using Eq.~(\ref{eq:macro_ks})  and compare it with $\chi_{\rm KS}(\vec q)$ from Eq.~(\ref{eq:delta_KS}) to show that the macroscopic static KS response function $\chi_{\rm KS}(\vec q)$ from the direct perturbation approach is in agreement with the LR-TDDFT result. 


\subsection{LR-TDDFT calculations of the macroscopic density response function with a consistent static XC kernel\label{sec:static_approximation}}

The workflow of the LR-TDDFT requires KS-orbitals from converged equilibrium KS-DFT simulation. These orbitals are used in the Dyson type equation (\ref{eq:Dyson}) together with the microscopic XC kernel $ K^{\rm xc}_{\scriptscriptstyle \vec G_1,\vec G_2}(\vec q, \omega)$ to compute the density response function, with  the microscopic XC kernel being the first order functional derivative of the XC potential. Consistent LR-TDDFT simulations require a microscopic XC kernel that is fully compatible with the XC functional used to compute the XC potential. 
However, the LR-TDDFT for extended systems (using the first order functional derivative of the XC potential to find the XC kernel) currently has the limitation that consistent simulations can be performed using only adiabatic LDA and GGA microscopic XC kernels ~\cite{Byun_2020}.
For a macroscopic density response function and dielectric function, we can circumvent this limitation for on-average-homogeneous systems (e.g. WDM, dense plasmas, and liquid metals) by combining the static macroscopic XC kernel $K_\textnormal{xc}(\mathbf{q})$ from the direct perturbation approach, Eq.~(\ref{eq:kernel}), with the dynamic macroscopic KS response function $\chi_{\rm KS}(\vec q,\omega)$ defined by Eq.~(\ref{eq:macro_ks}). 

Our \textit{adiabatic (static) approximation} for the macroscopic density response function reads:
\begin{eqnarray}\label{eq:kernel}\label{eq:chi_adiabatic}
 \chi(\mathbf{q},\omega) = \frac{\chi_{\rm KS}(\mathbf{q},\omega)}{1 -\left[v(q)+K_{\rm xc}(\mathbf{q})\right]\chi_{\rm KS}(\mathbf{q},\omega)},
\end{eqnarray}

 \begin{figure}\centering
\includegraphics[width=0.35\textwidth]{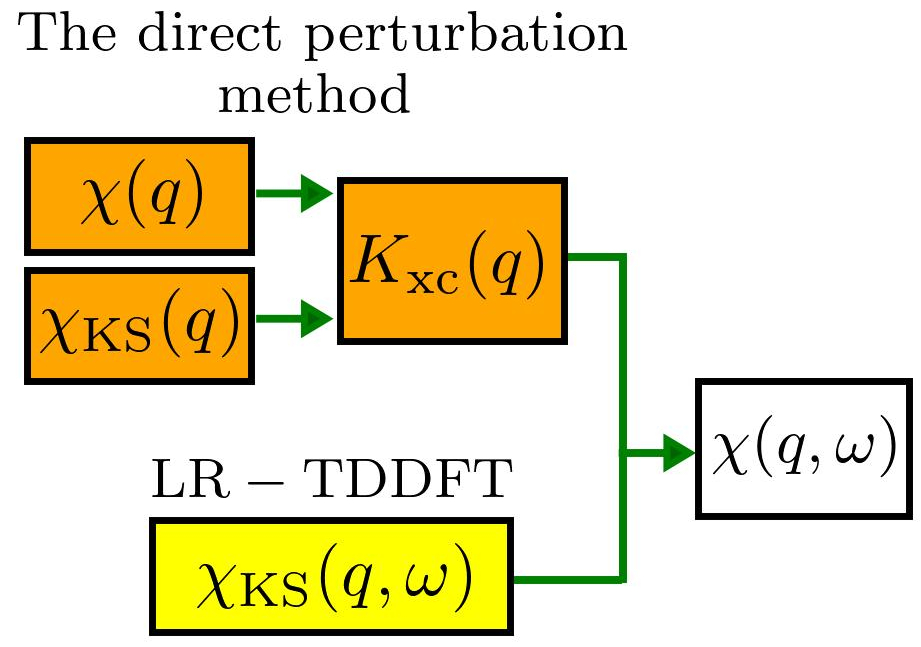}
\caption{\label{fig:scheme} Computation scheme for the macroscopic density response function Eq.~(\ref{eq:chi_adiabatic}) based on  the adiabatic (static) XC kernel from the direct perturbation approach Eq.~(\ref{eq:invert}) and the macroscopic KS response function Eq.~(\ref{eq:macro_ks}) from the LR-TDDFT. The presented self-consistent  calculation scheme can be used beyond ALDA with any XC functional across Jacob's ladder. 
}
\end{figure} 

The self-consistent  calculation method represented by Eq.~(\ref{eq:chi_adiabatic}) is schematically explained in Fig. \ref{fig:scheme}.
To deliver a macroscopic dynamic density response function in an adiabatic approximation, this approach combines the static XC kernel from the direct perturbation approach defined by Eq.~(\ref{eq:kernel}) with the macroscopic KS response function from the LR-TDDFT introduced by Eq.~(\ref{eq:macro_ks}). 
We reiterate that neither Eq.~(\ref{eq:macro_ks}) for $\chi_{\rm KS}(\vec q,\omega)$  nor  Eq.~(\ref{eq:invert}) for $K_\textnormal{xc}(\mathbf{q})$  require information about the microscopic XC kernel $ K^{\rm xc}_{\scriptscriptstyle \vec G_1,\vec G_2}(\vec q, \omega)$. In this way, we can compute the   macroscopic density response function $ \chi(\vec q,\omega)$ in an adiabatic approximation for any available XC functional in a fully self-consistent  manner.  The macroscopic dynamic dielectric function follows from  Eq.~(\ref{eq:df_plasma}).

Importantly, if a consistent adiabatic XC-kernel is used, a standard LR-TDDFT result for $\varepsilon_M(\vec q, \omega)$ as defined by Eq.~(\ref{eq:d_f_lf}) is equivalent to the result from our adiabatic approximation for the macroscopic density response function Eq.~(\ref{eq:chi_adiabatic}). We demonstrate it for the ALDA for the example of warm dense hydrogen in Sec. \ref{s:tddft_vs_dp}.

For real materials, the adiabatic approximation represented by Eq.~(\ref{eq:chi_adiabatic}) is more powerful than ALDA or AGGA of the standard LR-TDDFT since Eq.~(\ref{eq:chi_adiabatic}) can be computed using any XC functional across Jacob's ladder of DFT \cite{Jacob1, Jacob2}. For example, Moldabekov \textit{et al} \cite{Moldabekov_dft_kernel} have used the direct perturbation approach for the investigation of the static density response funtion and  XC kernel of the uniform electron gas and warm dense hydrogen using various LDA, GGA, and meta-GGA functionals. Additionally, different hybrid XC functionals have been analyzed at WDM conditions using the UEG model in Refs.~\cite{Moldabekov_non_empirical_hybrid, hybrid_results}. These prior works used the reference static density response function $\chi_0(q)$ from KS-DFT calculations with zero XC functional. This has been done for benchmarking purposes against available quantum Monte Carlo results. 
The novelty of Eq.~(\ref{eq:chi_adiabatic}) is the consistent way to compute $\chi_{0}(\vec q,\omega)=\chi_{\rm KS}(\vec q,\omega)$ and $K_{\rm xc}(q)$ for applications within the adiabatic (static) approximation.

For WDM and dense plasma applications, Eq.~(\ref{eq:chi_adiabatic}) does not require the introduction of an effective free electron density (in contrast to Lindhard function based approaches) and our static XC kernel is now material specific. This also eliminates the need of using some model local field correction in Eq.~(\ref{eq:LFC}).

We note that for the uniform electron gas (UEG), $\chi_{\rm KS}(\vec q,\omega)$ reduces to the Lindhard function as we show in  Sec. \ref{s:tddft_vs_dp}. In this case, quantum Monte Carlo simulations have shown that adiabatic (static) approximation is  highly accurate in the case of weak to moderate coupling strengths, e.g., at metallic densities~\cite{dornheim_dynamic}.

\section{ Simulation parameters}\label{s:smd}

We use Hartree atomic units for all numerical results presented in this work.
The total density of electrons is given in terms of the density parameter $r_s=\left(4\pi n/3\right)^{-1/3}$, which is the mean-inter particle distance.
We note that $r_s$ also represents the characteristic coupling parameter between electrons \cite{review, zhandos1, Moldabekov_PRE_15}.
The temperature of electrons is expressed in terms of the degeneracy parameter $\theta=T/T_F$, which is the temperature value in the units of the Fermi temperature (energy). 
We consider $r_s=2$ and $r_s=4$, which are typical for WDM experiments. For warm dense hydrogen we set $\theta=1$. This allows us to compare $\chi (q)$  from the KS-DFT simulations  with the recent exact path integral quantum Monte Carlo results by B\"ohme \textit{et al.}~\cite{Bohme_PRL_2022}. 

In order to test the validity of the presented approach for $\chi_{\rm KS} (q)$  in the limit of the UEG,  we also use $\theta=1$, $0.5$, and $0.01$. 

We set $\vec q$ along the z-axis and, having this in mind, we drop vector notation.   
In the case of the direct perturbation approach, the perturbation wave numbers are given by $q=j\times q_{\rm min}$ with $q_{\rm min}=2\pi/L$ and $j$ being positive integer numbers.
For a given snapshot of ionic positions, the density and KS potential perturbations were averaged along the $x$ and $y$ axes (since $\vec q$ is along the z-axis).
The size of the main simulation box is defined by $n_0 L^3=N$, where $n_0=3/(4\pi r_s^3)$ and $N$ is the total number of electrons in the main simulation box. We consider $N=14$ and $N=20$. At the considered parameters, the finite size effects at $N=14$ and $N=20$  are negligible. This was demonstrated by path integral quantum Monte Carlo simulations in Refs. \cite{Bohme_PRL_2022, dornheim_ML, dornheim_HEDP,  Dornheim_PRR_2021} and by KS-DFT simulations in Refs. \cite{Moldabekov_JCP_2021, Moldabekov_dft_kernel, moldabekov2021thermal, hybrid_results}.

In the case of the LR-TDDFT calculations, $q$ values must be the difference between two k-points. For example, if the \textit{k}-point grid is $N_k\times N_k \times N_k$, the wave number is given by $\vec q=\left(l/N_k, m/N_k, p/N_k\right)2\pi/L$, where $l,~m,~p$ are positive integer numbers and are not multiples of  $N_k$. Since we set $\vec q$ along z-axis, we always have $l=0$ and $m=0$.

The external field amplitude in the direct perturbation approach must be small enough so that the density perturbation can be described by the LRT. 
For warm dense hydrogen, we set $A=0.01$ (in Hartree). It was shown to be within the LRT domain in Ref. \cite{Bohme_PRL_2022}.
Following Ref. \cite{Moldabekov_JCTC_2022}, for the UEG calculations, we set $A=0.01$ at $r_s=2$ and $A=0.002$ at $r_s=4$.

For the KS-DFT calculations of the static XC kernel we use the ABINIT package \cite{Gonze2020, Romero2020, Gonze2016, Gonze2009, Gonze2005, Gonze2002} with the pseudopotentials by Goedecker, Teter and Hutter \cite{PhysRevB.54.1703}.
In case of the LDA functional \cite{Perdew_Wang_PRB_1992}, we cross checked that the ABINIT results are reproduced by the KS-DFT calculations using the GPAW code~\cite{GPAW1, GPAW2, ase-paper, ase-paper2}, which is a real-space implementation of the projector augmented-wave method.
For the LR-TDDFT calculations on the basis of the LDA, we used the GPAW code for the calculation of the density response function and KS response function \cite{LRT_GPAW2}.
The ionic configurations have been obtained from thermal KS-DFT based molecular dynamics simulations as it is described in Ref. \cite{Fiedler_PRR_2022}.

The direct perturbation approach based KS-DFT simulations of the warm dense hydrogen were performed for $14$ (with $280$ bands) and $20$ (with $400$ bands) particles in the main simulation cell.  We used $10\times10\times10$ k-points sampling and $30~{\rm Ha}$ energy cutoff.  For the LR-TDDFT calculations of the static density response functions  at $r_s=2$ and  $q<2q_F$, we used $14$ electrons in the main cell with $280$ bands and $500~{\rm eV}$ energy cutoff in the equilibrium state calculations. For  $q>2q_F$, we used the equilibrium state calculations with $1500$ bands for $14$ electrons in the main cell, $4\times4\times4$ k-points, and $550~{\rm eV}$ energy cutoff. Using wave functions from the equilibrium state, the LR-TDDFT calculations  were performed with the cutoff in the dielectric function set to $100~{\rm eV}$ and the broadening parameter $\eta=0.001$.

 The  calculations for the static density response functions of the UEG at $\theta=0.01$ and $\theta=0.5$ were performed using $N=38$ particles in the main cell (with $44$ bands), $10\times10\times10$ k-points sampling,  and $30~{\rm Ha}$ energy cutoff. The data for $\theta=1$ is obtained using $N=20$ electrons in the main cell (with $400$ bands), $10\times10\times10$ k-points sampling, and $30~{\rm Ha}$ energy cutoff.

\section{Simulation results}\label{s:all_results}

 \begin{figure}\centering
\includegraphics[width=0.45\textwidth]{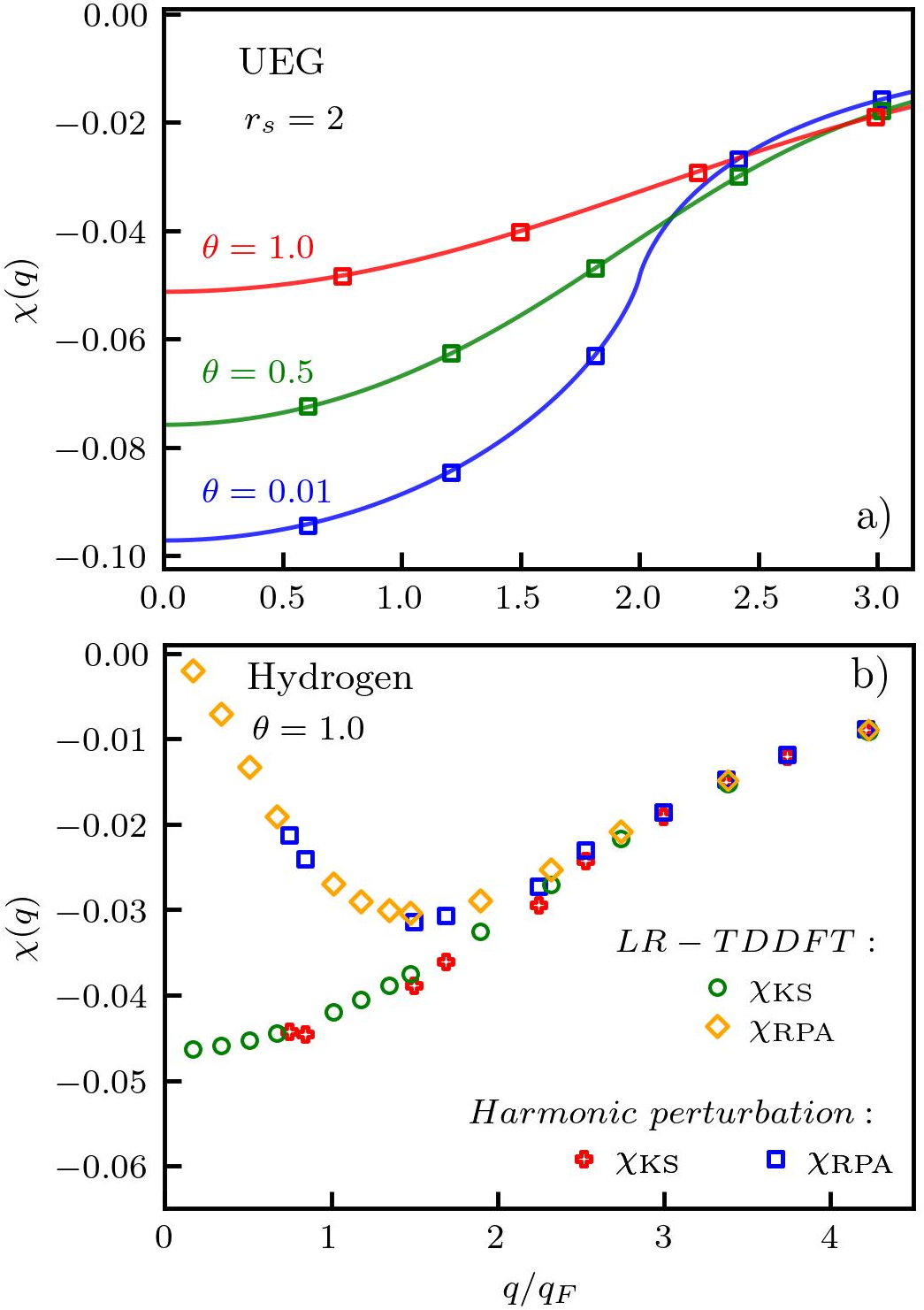}
\caption{\label{fig:ks_chi_rs2} a) Symbols show KS response function computed from the KS-DFT simulations of the UEG with LDA XC functional for  $r_s=2$ at different values of the reduced temperature $\theta$.  Solid lines are corresponding results computed using the  Lindhard function. b) KS response function and corresponding RPA values for warm dense hydrogen at $r_s=2$ and $\theta=1$ as computed using the LR-TDDFT (green and orange symbols) and from the direct perturbation technique (red and blue symbols). 
}
\end{figure} 

 \begin{figure}[t!]\centering
\includegraphics[width=0.45\textwidth]{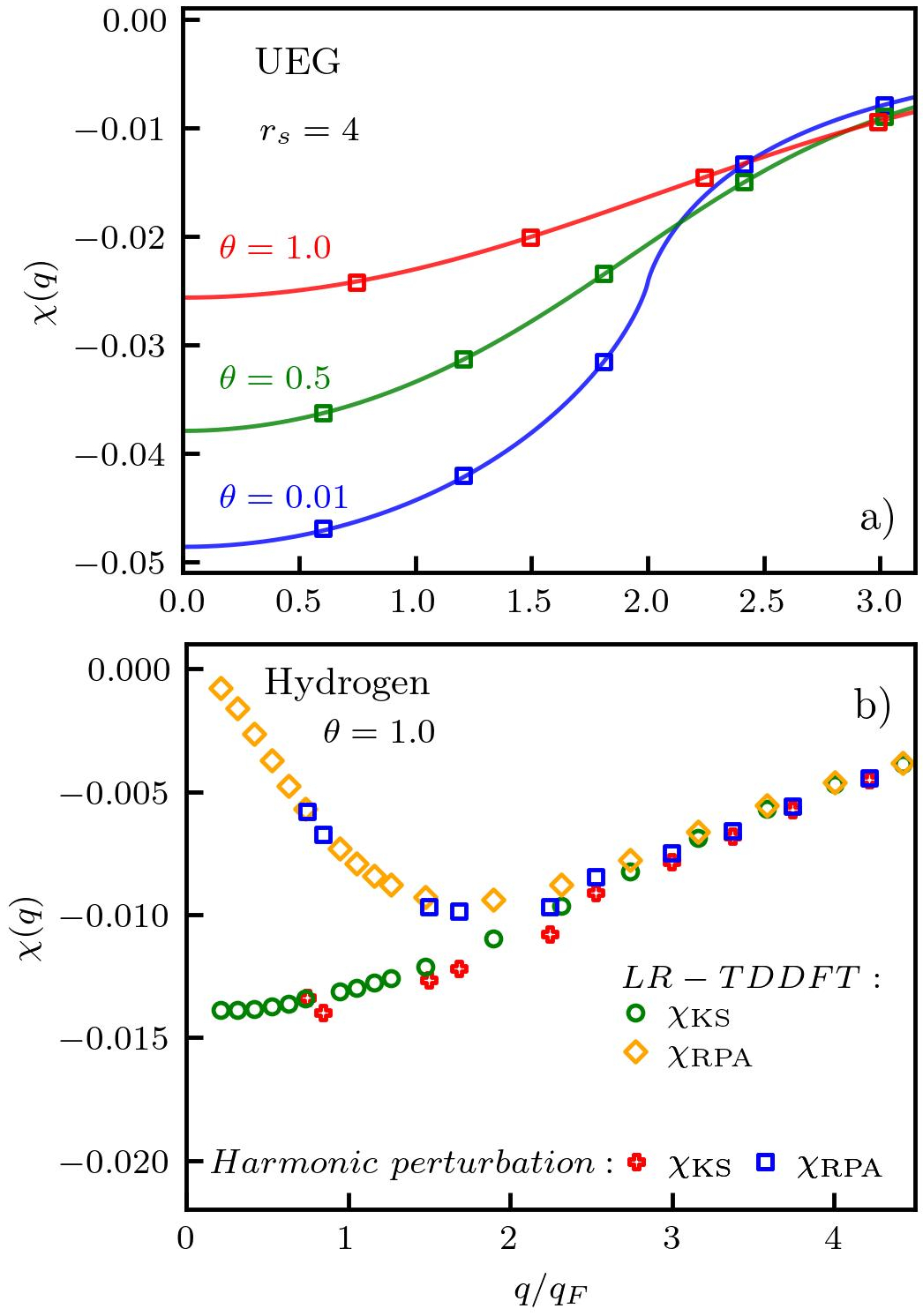}
\caption{\label{fig:ks_chi_rs4} a) Symbols show KS response function computed from the KS-DFT simulations of the UEG with LDA XC functional for  $r_s=4$ at different values of the reduced temperature $\theta$.  Solid lines are corresponding results computed using the  Lindhard function. b) KS response function and corresponding RPA values for warm dense hydrogen at $r_s=4$ and $\theta=1$ as computed using the LR-TDDFT (green and orange symbols) and from the direct perturbation technique (red and blue symbols). 
}
\end{figure} 

\subsection{Equivalence of LR-TDDFT and the direct perturbation based approach}\label{s:tddft_vs_dp}


To test the correctness of the macroscopic KS response function computed using Eq.~(\ref{eq:delta_KS}), first we consider the uniform electron gas~\cite{review} for which the macroscopic KS response function must be equivalent to the Lindhard function.  We show that it is indeed the case for $r_s=2$ in Fig. \ref{fig:ks_chi_rs2}a) and for $r_s=4$ in Fig. \ref{fig:ks_chi_rs4}a) with $\theta$ covering both the ground state and the WDM regime. 
From  Figs. \ref{fig:ks_chi_rs2}a) and \ref{fig:ks_chi_rs4}a), we see that  the macroscopic KS response function Eq.~(\ref{eq:delta_KS}) perfectly reproduces the Lindhard function in the thermodynamic limit for the UEG. To get the data presented in this figures we used the LDA functional by Perdew and Wang~\cite{LDA_PW}. Therefore, Eq.~(\ref{eq:delta_KS}) provides the correct non-interacting (ideal) response function even if the KS-DFT calculations are performed with non-zero XC functional.

Second, in Fig. \ref{fig:ks_chi_rs2}b) and Fig. \ref{fig:ks_chi_rs4}b), we demonstrate that the macroscopic static KS response function of warm dense hydrogen at $\theta=1$ computed using the direct perturbation approach Eq.~(\ref{eq:delta_KS}) is equivalent to the result found using the standard LR-TDDFT method via Eq.~(\ref{eq:macro_ks}).
Fig. \ref{fig:ks_chi_rs2}b) and Fig. \ref{fig:ks_chi_rs4}b) present results for $r_s=2$ and $r_s=4$, respectively, at $\theta=1$. Additionally, we show the results in the RPA, which are computed using $\chi_{\rm KS}(q)$ in Eq.~(\ref{eq:chi_p}) with $\omega=0$. From Figs. \ref{fig:ks_chi_rs2}b) and \ref{fig:ks_chi_rs4}b), one can see that the $\chi_{\rm KS}(q)$ computed using Eq.~(\ref{eq:delta_KS}) is in agreement with the  $\chi_{\rm KS}(q)$ obtained using Eq.~(\ref{eq:macro_ks}).

\begin{figure}\centering
\includegraphics[width=0.45\textwidth]{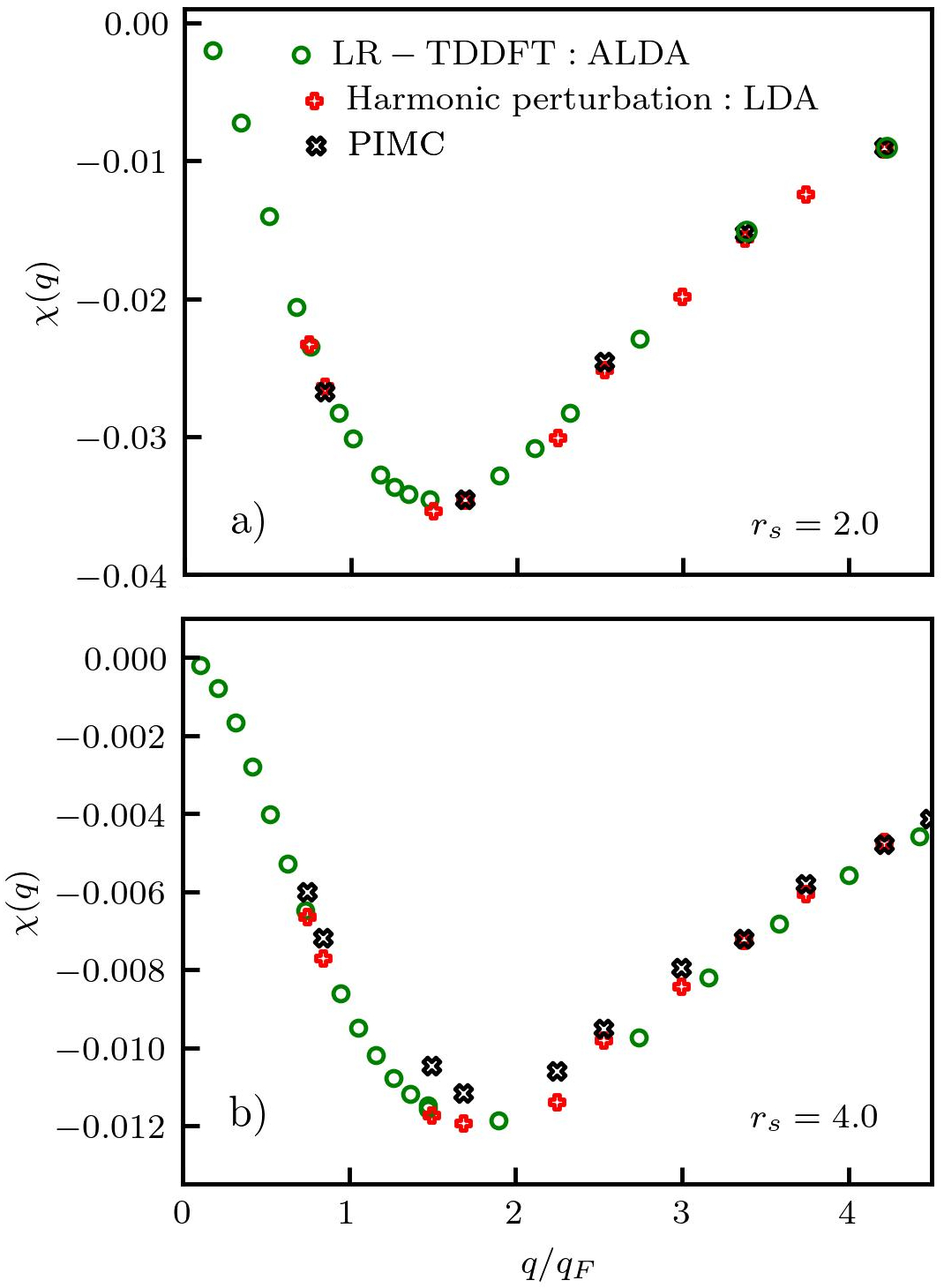}
\caption{\label{fig:chi_alda} Density response function of warm dense hydrogen at a) $r_s=2$ and b) $r_s=4$ for $\theta=1$. The results computed using the LR-TDDFT with ALDA kernel and the harmonic perturbation based approach using LDA functional are compared with the exact  PIMC  data from Ref. \cite{Bohme_PRL_2022}.  
}
\end{figure}

\begin{figure}[t]\centering
\includegraphics[width=0.45\textwidth]{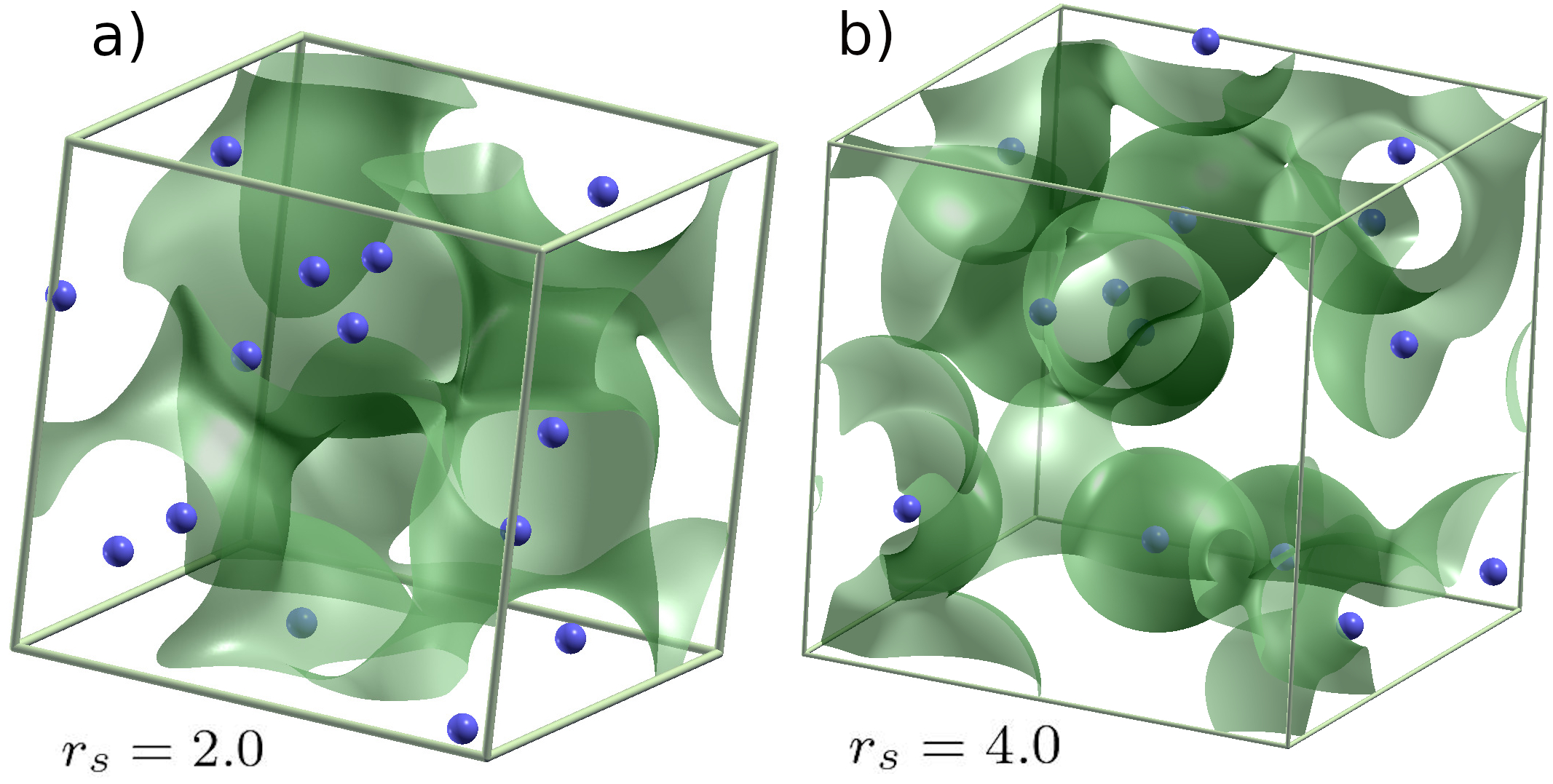}
\caption{\label{fig:illustration} Illustration of the stronger localization of the electrons around protons with the decrease in the density from a) $r_s=2$ to b) $r_s=4$ at fixed $\theta=1$ using an isosurface of the electronic density as obtained from DFT-MD. This figure was created using  XCrySDen \cite{XCrySDen1, XCrySDen2}.
}
\end{figure}

Next, we show in Fig. \ref{fig:chi_alda} that the density response function $\chi(q)$ computed according to Eq.~(\ref{eq:delta_KS}) within the direct perturbation approach using the LDA XC functional is in agreement with the $\chi(q)$ from the LR-TDDFT calculations using the ALDA kernel and equilibrium state wave functions generated using LDA. Fig. \ref{fig:chi_alda}a) shows the data for $r_s=2$ and  Fig. \ref{fig:chi_alda}b) presents the results for $r_s=4$.
Additionally, in Fig. \ref{fig:chi_alda}, we compare the KS-DFT results for $\chi(q)$ with the recent exact path-integral quantum Monte Carlo (PIMC) results \cite{Bohme_PRL_2022}.
From Fig. \ref{fig:chi_alda}a) one can see that the KS-DFT results are in close agreement with the PIMC data at $r_s=2$. In Ref. \cite{Bohme_PRL_2022}, it was shown that at $\theta=1$ and $r_s=2$, electrons in the warm dense hydrogen manifest free-electron like behavior indicating weak electron-ion coupling.
In contrast, at $r_s=4$ and $\theta=1$, electron-ion coupling is strong and the system can effectively be described as partially  ionized. 
As the result, in Fig. \ref{fig:chi_alda}b), we observe that the quality of the LDA based KS-DFT data around $2q_F$ significantly deteriorates compared to the exact PIMC data with the decrease in the density from $r_s=2$ to $r_s=4$. The deterioration of  the quality of the LDA based description with the decrease in density is due to the stronger localization of electrons around ions \cite{Bohme_PRE_2022}, which gives rise to self-interaction errors in KS-DFT \cite{doi:10.1126/science.1158722}. Stronger coupling between electrons and ions also leads to the smaller density response amplitude as it can be seen by comparing the $\chi(q)$ values in Fig. \ref{fig:chi_alda}a) and Fig. \ref{fig:chi_alda}b).
The stronger localization of electrons around ions at $r_s=4$ compared to the case with $r_s=2$ in warm dense hydrogen is illustrated using an isosurface of the electronic density in Fig. \ref{fig:illustration}.

 \begin{figure*}\centering
\includegraphics[width=0.85\textwidth]{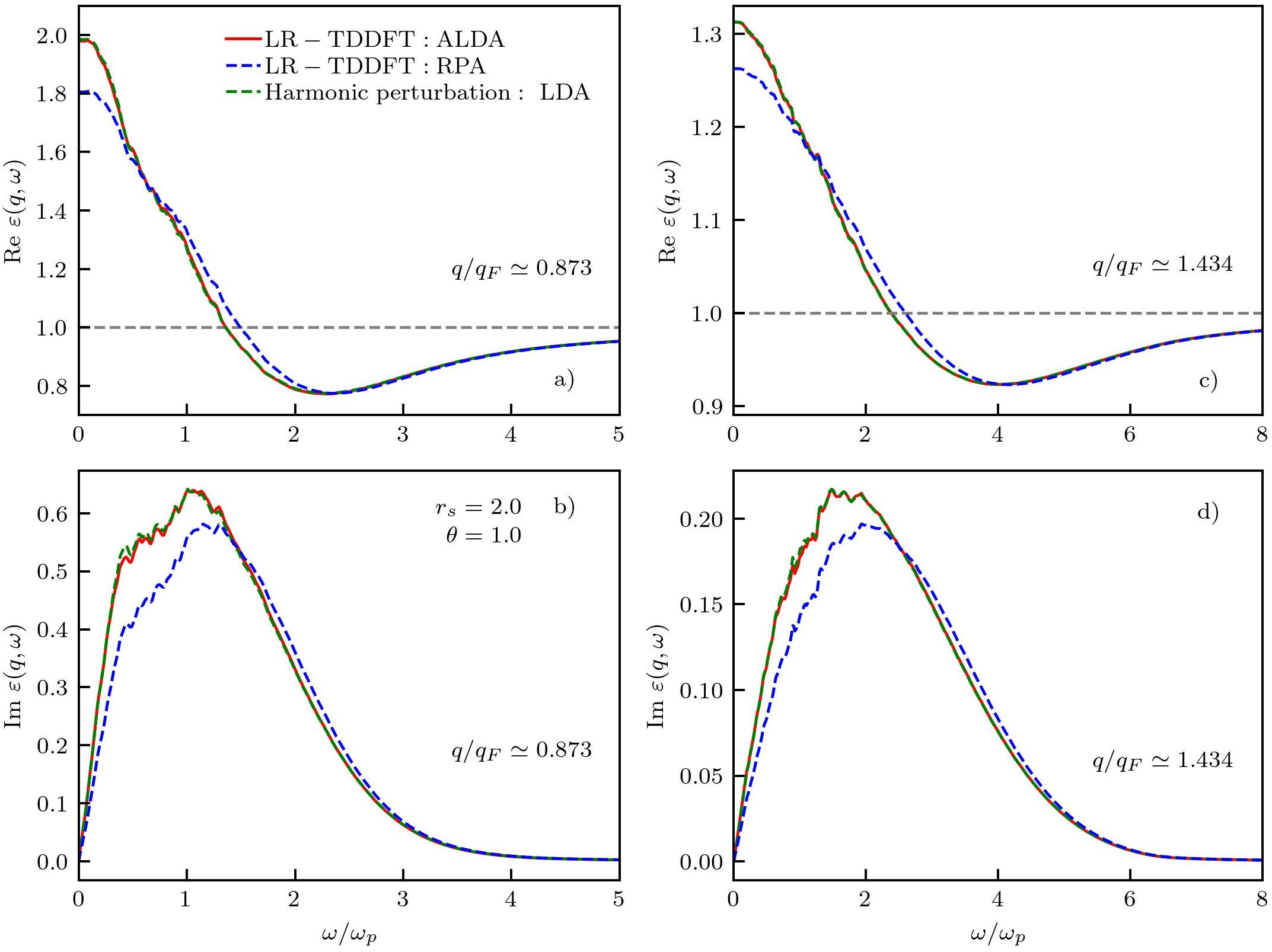}
\caption{\label{fig:df} Demonstration of the equivalence of the LR-TDDFT result with the ALDA kernel (solid red line) to the macroscopic XC kernel and macroscopic KS response function based scheme (as explained in Fig. \ref{fig:scheme}) using the LDA functional (dashed green). The results are presented  for warm dense hydrogen at  $r_s=2$ and $\theta=1$.  a) The real part of the dynamic dielectric function and b) the imaginary part of the dynamic dielectric function at $q/q_F\simeq 0.873$. c) The real part of the dynamic dielectric function and d) the imaginary part of the dynamic dielectric function at $q/q_F\simeq 1.434$. The RPA result is provided for comparison as well (dashed blue).
}
\end{figure*}

The third quantity we consider is the macroscopic  dynamic dielectric function $\varepsilon_M (q, \omega)$ of warm dense hydrogen.
We compute $\varepsilon_M (q, \omega)$  using the standard LR-TDDFT with ALDA kernel and LDA based KS orbitals. Additionally, 
we calculate $\varepsilon_M (q, \omega)$ using  $\chi(q,\omega)$ from Eq.~(\ref{eq:chi_adiabatic}) in Eq.~(\ref{eq:df_plasma}) according to the scheme presented in Fig. \ref{fig:scheme}, where $\chi_{\rm KS}(q,\omega)$  is computed using Eq.~(\ref{eq:macro_ks}) and the LDA XC kernel is obtained from the direct perturbation approach using Eq.~(\ref{eq:invert}). The comparison of the data for $\varepsilon_M (q, \omega)$ computed using these two different ways is presented in  Fig. \ref{fig:df}, where we  show the real and imaginary parts of  $\varepsilon_M (q, \omega)$  at $q\simeq  0.873 q_F$ and $q\simeq 1.434 q_F$. Additionally, in Fig. \ref{fig:df}, we present the RPA results computed using zero XC kernel. From Fig. \ref{fig:df}, we observe that the ALDA based LR-TDDFT data  for $\varepsilon_M (q, \omega)$  is in excellent agreement with the result obtained using $\chi(q,\omega)$ from Eq.~(\ref{eq:chi_adiabatic}) in Eq.~(\ref{eq:df_plasma}).

Therefore, for the example of the warm dense hydrogen, we have demonstrated using the LDA functional that our approach from Eq.~(\ref{eq:chi_adiabatic}) for the macroscopic $\chi(q,\omega)$ is equivalent to  the standard LR-TDDFT  with an adiabatic XC kernel approximation. We reiterate that the strength of the presented new approach is that it can be used with the static XC kernel of any available XC  functional without explicitly computing the second order functional derivative. For example, using the direct perturbation approach, Moldabekov \textit{et al.} \cite{Moldabekov_dft_kernel} presented a static XC kernel for LDA, GGA, and meta-GGA level functionals as well as for various hybrid functionals \cite{hybrid_results, Moldabekov_non_empirical_hybrid}.    
Moreover, Eq.~(\ref{eq:chi_adiabatic})  represents a method where the adiabatic XC kernel $K_{\rm XC} (q)$ is fully consistent to the XC functional used to generate the  KS response function $\chi_{\rm KS}(q,\omega)$ (defined by Eq.~(\ref{eq:macro_ks})).

\subsection{Static XC kernel for warm dense hydrogen at metallic density}\label{s:rs2}

In Ref. \cite{Moldabekov_dft_kernel},  the PIMC data based comparative analysis  of the density response function of warm dense hydrogen at $r_s=2$ and $r_s=4$ using  PBE \cite{PBE}, PBEsol \cite{PBEsol}, AM05 \cite{PhysRevB.72.085108},  and SCAN \cite{SCAN} functionals revealed that these XC functionals do not improve the description of the warm dense hydrogen compared  to ground state LDA \cite{LDA_PW} based calculations. Moreover, it was shown that a finite-temperature LDA functional \cite{groth_prl} developed for WDM conditions does not improve the quality of the description compared to the ground state LDA and even provides slightly worse results. Among aforementioned functionals in Ref.~\cite{Moldabekov_dft_kernel},  the meta-GGA level SCAN functional occupies a higher rung on Jacob's ladder compared to the LDA and GGA functionals and is more accurate than the LDA and PBE for the ground state applications. Thus, it is surprising that SCAN performs worse than the ground state LDA and PBE when the thermal energy is comparable with the Fermi energy. 
It is a natural question to ask whether other meta-GGA functionals have the same drawback in the WDM regime. 
Thus, in this work, in addition to the ground state LDA and PBE, we included another two meta-GGA XC potentials into our analysis.
We consider XC potentials introduced by Tran and Blaha \cite{Tran} (TB) and by R\"as\"anen, Pittalis, and Proetto (RPP) \cite{Rasanen}.
These potentials  are introduced by modifying the Becke–Johnson (BJ) exchange potential \cite{Becke_Johnson} (approximating an exact Hartree-Fock exchange) and by adding  the LDA correlation by Perdew and Wang \cite{LDA_PW}. 
The TB approximation provides high quality results for various types of solids and a good agreement with experiments. This approximation recovers the ground sate LDA approximation for a constant electronic density. The RPP approximation is designed to be exact for any single-electron system and to reduce to the LDA level description in the limit of a constant density. For example, the RPP approximation proved to be very accurate for a hydrogen chain in an external electric field \cite{Rasanen}. 

 \begin{figure}\centering
\includegraphics[width=0.45\textwidth]{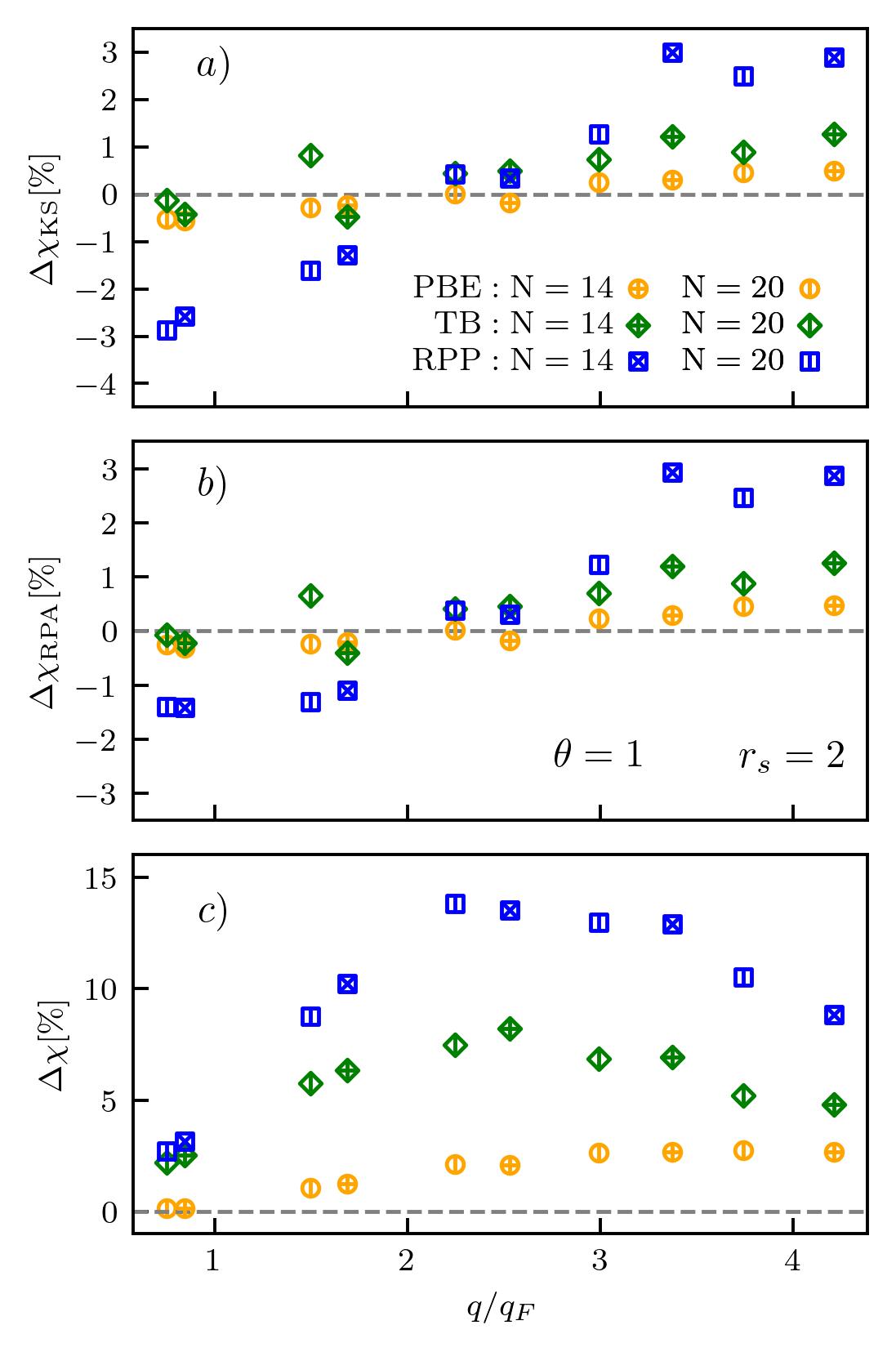}
\caption{\label{fig:ks_rs2} The relative deviation from the LDA results of a) the KS response function, b) the screened response on the level of RPA, and c) the  total density response function from simulations using GGA level PBE functional, meta-GGA level TB and RPP approximations for warm dense hydrogen at  $r_s=2$ and $\theta=1$. 
}
\end{figure}

We start our discussion of the results for warm dense hydrogen at a metallic density  $r_s=2$.
We consider the static KS response function $\chi_{\rm KS}(q)$, the static density response in the RPA $\chi_{\rm RPA}(q)$, and the static total density response function $\chi(q)$. 
In Fig. \ref{fig:ks_rs2}, we analyse the relative deviation of $\chi_{\rm KS}(q)$, $\chi_{\rm RPA}(q)$, and $\chi(q)$ from the LDA based data for $r_s=2$ and $\theta=1$.
The results are presented for $14$ and $20$ particles in the main cell. 
The relative deviation  from the LDA based data is computed as 
\begin{equation}\label{eq:diff_ks}
    \Delta \chi_{\rm KS}(q)[\%]=\frac{\chi_{\rm KS}(q)-\chi_{\rm KS}^{\rm LDA}(q)}{\chi_{\rm KS}^{\rm LDA}(q)}\times 100\%,
\end{equation}
where $\chi_{\rm KS}^{\rm LDA}(q)$ is the static KS response function calculated using the LDA.

Similar expressions are used to evaluate the relative deviations of  $\chi_{\rm RPA}(q)$ and $\chi(q)$  from corresponding LDA based results $\chi_{\rm RPA}^{\rm LDA}(q)$ and $\chi^{\rm LDA}(q)$:
\begin{equation}\label{eq:diff_rpa}
    \Delta \chi_{\rm RPA}(q)[\%]=\frac{\chi_{\rm RPA}(q)-\chi_{\rm RPA}^{\rm LDA}(q)}{\chi_{\rm RPA}^{\rm LDA}(q)}\times 100\%,
\end{equation}
and 
\begin{equation}\label{eq:diff_tot_chi}
    \Delta \chi(q)[\%]=\frac{\chi(q)-\chi^{\rm LDA}(q)}{\chi^{\rm LDA}(q)}\times 100\%.
\end{equation}

 \begin{figure}\centering
\includegraphics[width=0.45\textwidth]{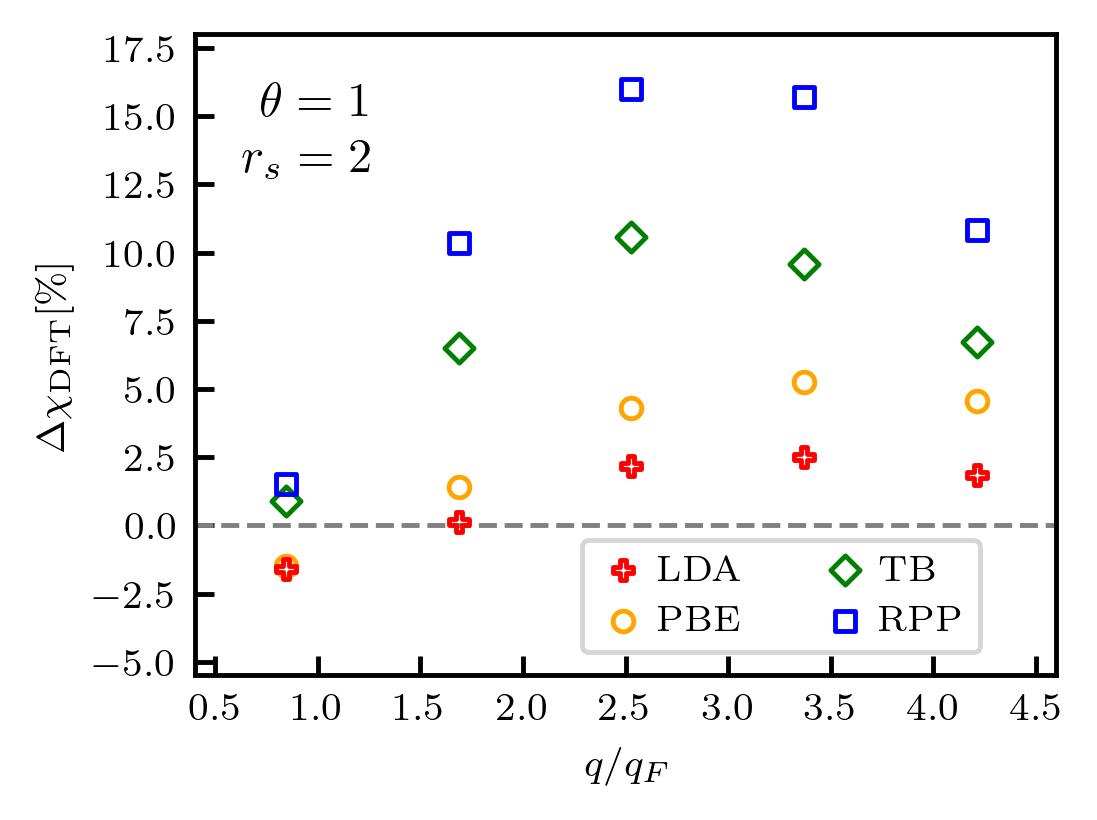}
\caption{\label{fig:chi_err_rs2} The deviation of the KS-DFT results from the exact PIMC data for the static density response function of warm dense hydrogen at $r_s=2$ and $\theta=1$. 
}
\end{figure} 

Fig. \ref{fig:ks_rs2}a) shows that the PBE and TB based data for  $\chi_{\rm KS}(q)$ are in very close agreement to the $\chi_{\rm KS}^{\rm LDA}(q)$ with the relative deviation 
smaller than $2\%$ by absolute value. The RPP based results for   $\chi_{\rm KS}(q)$ have slightly larger disagreement with $\chi_{\rm KS}^{\rm LDA}(q)$, which is in between $3\%$ and $-3\%$. These  disagreements are further diminished at $q<2q_F$ due to screening as one can see from Fig. \ref{fig:ks_rs2}b), where the results for  the $\Delta \chi_{\rm RPA}(q)$ are presented. In $\chi_{\rm RPA}(q)$, screening is taken into account using Eq.~(\ref{eq:chi_p}), which neglects the XC kernel. Overall, we see that the results for  $\chi_{\rm KS}(q)$ as well as  $\chi_{\rm RPA}(q)$ computed using different considered  XC potentials are close to  each other. In fact, at $r_s=2$ and $\theta=1$,  $\chi_{\rm KS}(q)$ has also close values to the Lindhard function derived using the ideal electron gas model (see Fig. \ref{fig:ks_chi_rs2}). This is in agreement with previously discussed observation that at $r_s=2$ and $\theta=1$ electrons are strongly delocalised.

\begin{figure}\centering
\includegraphics[width=0.45\textwidth]{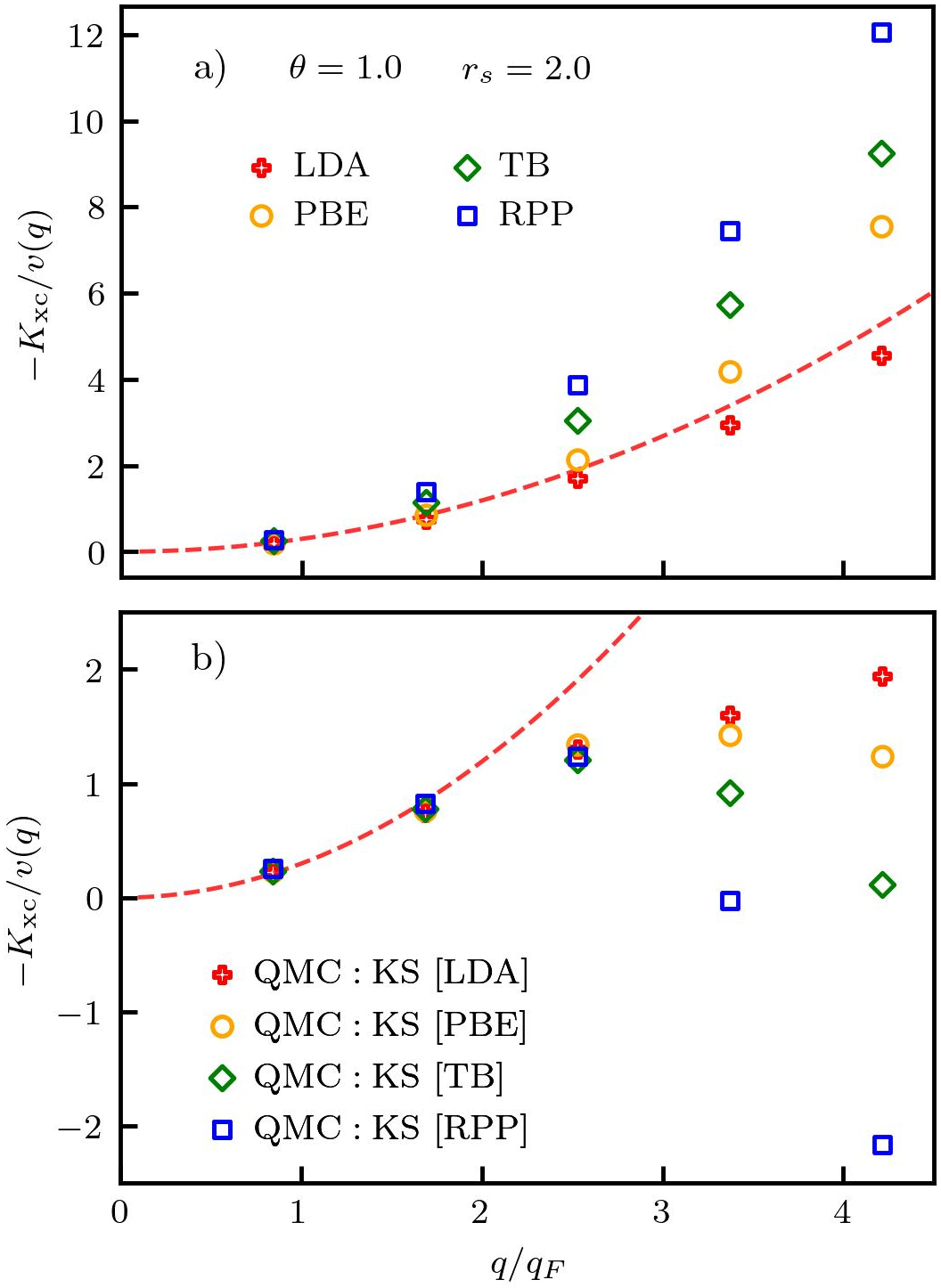}
\caption{\label{fig:LFCrs2} a) Fully consistent static XC-kernel from KS-DFT simulations with different XC functionals. b) Static XC-kernel extracted using the PIMC data and KS-response function from KS-DFT simulations. Demonstration of the equivalence of the KS response function from the harmonic perturbation technique and  LR-TDDFT calculations.  The results are for warm dense hydrogen at $r_s=2$ and $\theta=1$. 
}
\end{figure} 

The results for the actual density response $\chi(q)$  of warm dense hydrogen (cf. Eq.~(\ref{eq:delta_n})) to an external field are presented in Fig. \ref{fig:ks_rs2}c).
From Fig. \ref{fig:ks_rs2}c), we see that the $\chi(q)$ computed using the TB and RPP approximations significantly differs from the LDA based data at $q>q_F$. It is also the case for the PBE based data for $q>2q_F$. This is in contrast to the behavior of  $\chi_{\rm KS}(q)$ and  $\chi_{\rm RPA}(q)$, for which considered XC potentials give nearly the same results. 
Another important observation from Fig. \ref{fig:ks_rs2} is that the results computed using 14 and 20 particles exhibit a very similar behavior.
This means that the results computed at  $r_s=2$ and $\theta=1$ using 14 particles in the main cell are not affected by finite size effects.
This further corroborates previous results presented in Refs.~\cite{Moldabekov_dft_kernel, Bohme_PRL_2022}.

In Fig. \ref{fig:chi_err_rs2}, we analyze the quality of the considered approximations by comparing to the exact PIMC data from Ref.~\cite{Bohme_PRL_2022} for 14 particles in the main cell.
The relative deviation  of the KS-DFT data from the PIMC results is computed as 
\begin{equation}\label{eq:dft_vs_pimc}
    \Delta \chi_{\rm DFT}(q)[\%]=\frac{\chi(q)-\chi^{\rm PIMC}(q)}{\chi^{\rm PIMC}(q)}\times 100\%,
\end{equation}
where $\chi^{\rm PIMC}(q)$ is the density response function from the PIMC simulations using  the direct perturbation approach \cite{Bohme_PRL_2022}.

From Fig. \ref{fig:chi_err_rs2}, it is clear that the ground state LDA provides more accurate data for $\chi(q)$ compared to the PBE, TB, and RPP.

In Fig. \ref{fig:LFCrs2}, we present the corresponding results for the static XC kernel. In Fig. \ref{fig:LFCrs2}a), we show the static XC kernel computed self-consistently from the KS-DFT according to Eq.~(\ref{eq:invert}). In Fig. \ref{fig:LFCrs2}b), we present the static XC kernel obtained using the exact  $\chi(q)$ from the PIMC in Eq.~(\ref{eq:invert}). For both cases, the $\chi_{\rm KS}(q)$ is found from the KS-DFT simulations according to Eq.~(\ref{eq:delta_KS}). The dashed red line in Fig. \ref{fig:LFCrs2} represents a quadratic dependence of the LFC $G(q)$ in Eq.~(\ref{eq:LFC}) (i.e., the long-wavelength approximation) according to the compressibility sum-rule \cite{Moldabekov_dft_kernel}. 

From Fig. \ref{fig:LFCrs2}a), one can see that the PBE, TB, and RPP results for the XC kernel converge to the LDA based result at small wave numbers $q<2q_F$ and deviate with the increase in the wave number at $q>2q_F$. At considered wave numbers, the LDA based data closely follows the quadratic dependence. 
This is also the case for the PIMC based XC kernel at $q<2q_F$ as one can see in Fig. \ref{fig:LFCrs2}b).
We observe that the XC kernels computed  at $q<2q_F$ using the LDA, PBE, TB, and RPP based $\chi_{\rm KS}(q)$ are in a perfect agreement with the long-wavelength approximation, i.e., they have  a quadratic dependence. However, these XC kernels show a strong deviation from the quadratic dependence towards lower values with the increase in the wave number at $q>2q_F$.
Furthermore, if one compares the position of the PBE, TB, and RPP based results relative to the dashed red line in Fig. \ref{fig:LFCrs2}a) and Fig. \ref{fig:LFCrs2}b), we see that the corresponding XC kernel values deviate in different directions from the quadratic dependence in Fig. \ref{fig:LFCrs2}a) and Fig. \ref{fig:LFCrs2}b).
Overall, from comparing results in Fig. \ref{fig:LFCrs2}a) and Fig. \ref{fig:LFCrs2}b), we conclude that the LDA, PBE, TB, and RPP  give accurate results for the XC kernel at $q<2q_F$.
At $q>2q_F$,  the  PBE, TB, and RPP  fail to describe the XC kernel not only quantitatively, but also qualitatively. 

\begin{figure}\centering
\includegraphics[width=0.45\textwidth]{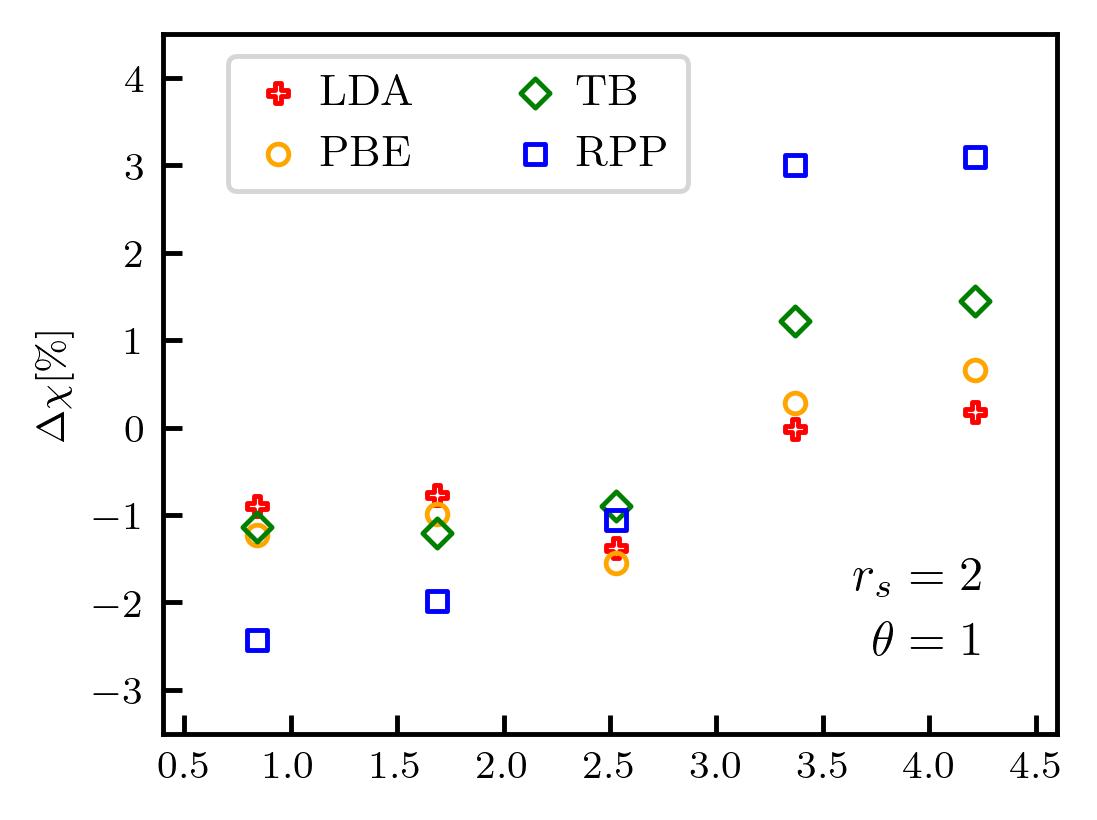}
\caption{\label{fig:NullXCrs2} Inaccuracy in the density response function due to an inconsistent combination of the KS response function and the XC kernel extracted from the PIMC data and an ideal reference function without XC effects. The results are for warm dense hydrogen at $r_s=2$ and $\theta=1$. 
}
\end{figure}

Another interesting question is whether it is possible to devise an universal XC kernel that can be used in the LR-TDDFT in combination with $\chi_{\rm KS}(q)$ to closely reproduce exact data for the density response properties in a wide range of parameters (temperatures, densities, and wave numbers). Our analysis of the static density response function and XC kernel indicates that there is no such  universal XC kernel for real materials.
One of many reasons could be that $\chi_{\rm KS}(q)$ depends on the used approximation for the XC potential. We observe it to a lesser extent at $r_s=2$ and to a greater extent at $r_s=4$ (as we show in Sec. \ref{s:rs4}), but it is always the case. In other words, the XC effects are included into both $\chi_{\rm KS}(q)$ and $K_{\rm xc}(q)$. 
To mitigate this effect, we computed  $\chi_{\rm KS}(q)$ with the XC functional set to zero. We denote the corresponding KS response function as  $\chi_{\rm KS}^{\rm NXC}(q)$ (with NXC standing for ``null XC''). We note that in this case we have non-interacting  electrons in an external field of ions.
Then we use $\chi_{\rm KS}^{\rm NXC}(q)$ in combination with the exact PIMC data for $\chi(q)$ in Eq.~(\ref{eq:kernel}) to compute  $K_{\rm xc}(q)$.
The computed XC kernel is then combined with the $\chi_{\rm KS}(q)$ computed using non-zero XC functional (the LDA, PBE, TB, and RPP) to find the density response function  $\chi(q)$ using Eq.~(\ref{eq:chi_adiabatic}). From Fig. \ref{fig:NullXCrs2} we see that the deviation of the $\chi(q)$  computed in this way from the exact  PIMC data is a few percent, but not zero. Thus, it does not reproduce the exact PIMC results for $\chi(q)$. However, we note that the error in  the $\chi(q)$ in Fig. \ref{fig:NullXCrs2} is significantly smaller  than deviations  shown in Fig. \ref{fig:chi_err_rs2}. This indicates that at $r_s=2$ and $\theta=1$, the $\chi_{\rm KS}^{\rm NXC}(q)$  is a good universal reference function for the computation of the XC kernel using the exact data for $\chi(q)$. We connect this finding to the fact that at $r_s=2$ and $\theta=1$, the warm dense hydrogen is nearly fully ionized and electron properties are similar to that of free electron gas. Indeed, we show in Sec. \ref{s:rs4}  that strong electronic localization around ions at $r_s=4$ leads to a significant worsening  of the  $\chi_{\rm KS}^{\rm NXC}(q)$   based approach.

 \begin{figure}\centering
\includegraphics[width=0.45\textwidth]{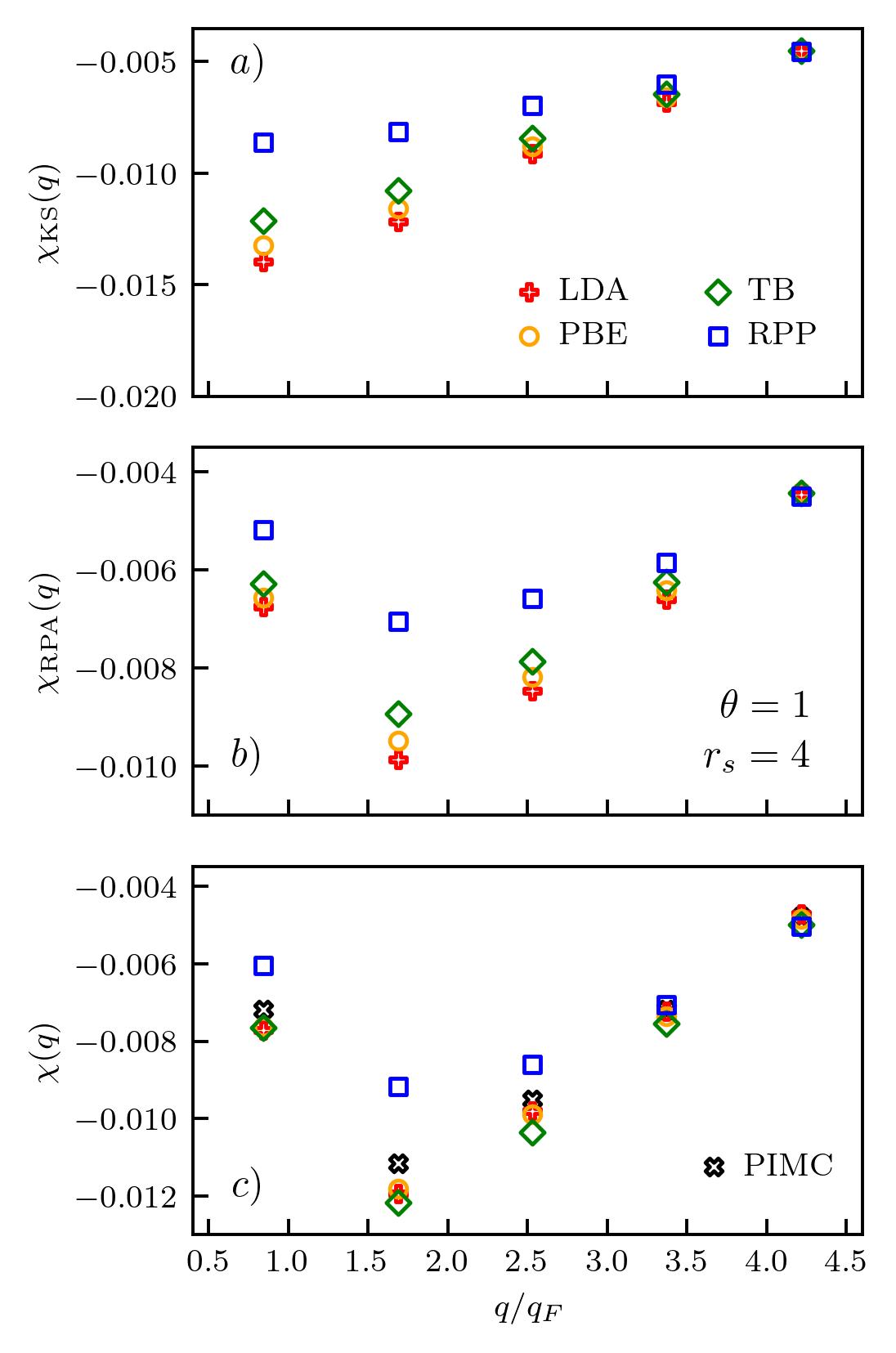}
\caption{\label{fig:chi_rs4} a) The KS response function, b) the screened response on the level of RPA, and c) the  total density response function from simulations using LDA,  PBE, and meta-GGA level TB and RPP approximations for warm dense hydrogen at  $r_s=4$ and $\theta=1$. 
}
\end{figure}


\subsection{Static XC kernel for strongly correlated warm dense hydrogen}\label{s:rs4}

Now we consider warm dense hydrogen at $r_s=4$ and $\theta=1$. At these parameters, we have a stronger electron-electron as well as electron-ion coupling compared to the case with $r_s=2$ and $\theta=1$. The results for $\chi_{\rm KS}(q)$, $\chi_{\rm RPA}(q)$, and $\chi(q)$  are presented in Fig. \ref{fig:chi_rs4} (with 14 particles in the main cell). 
From  Fig. \ref{fig:chi_rs4}a), one can see that there are strong disagreements  between $\chi_{\rm KS}(q)$ values computed using the LDA, PBE, TB, and RPP at $q<3q_F$. 
For $\chi_{\rm RPA}(q)$ these disagreements are somewhat alleviated due to screening as we see from  Fig. \ref{fig:chi_rs4}b). 
The results for the total density response function $\chi(q)$ are presented in Fig. \ref{fig:chi_rs4}c), where we provide a comparison with the exact PIMC data. 
From Fig. \ref{fig:chi_rs4}c), we see the LDA and PBE based data have similar quality compared to the exact PIMC data. The TB based data is slightly worse  than LDA and PBE.
The  results computed using RPP approximation show largest disagreement with the PIMC data. Overall,  the presented KS-DFT results have largest deviation from the PIMC data around $1.5 q_F<q< 2 q$.

\begin{figure}\centering
\includegraphics[width=0.45\textwidth]{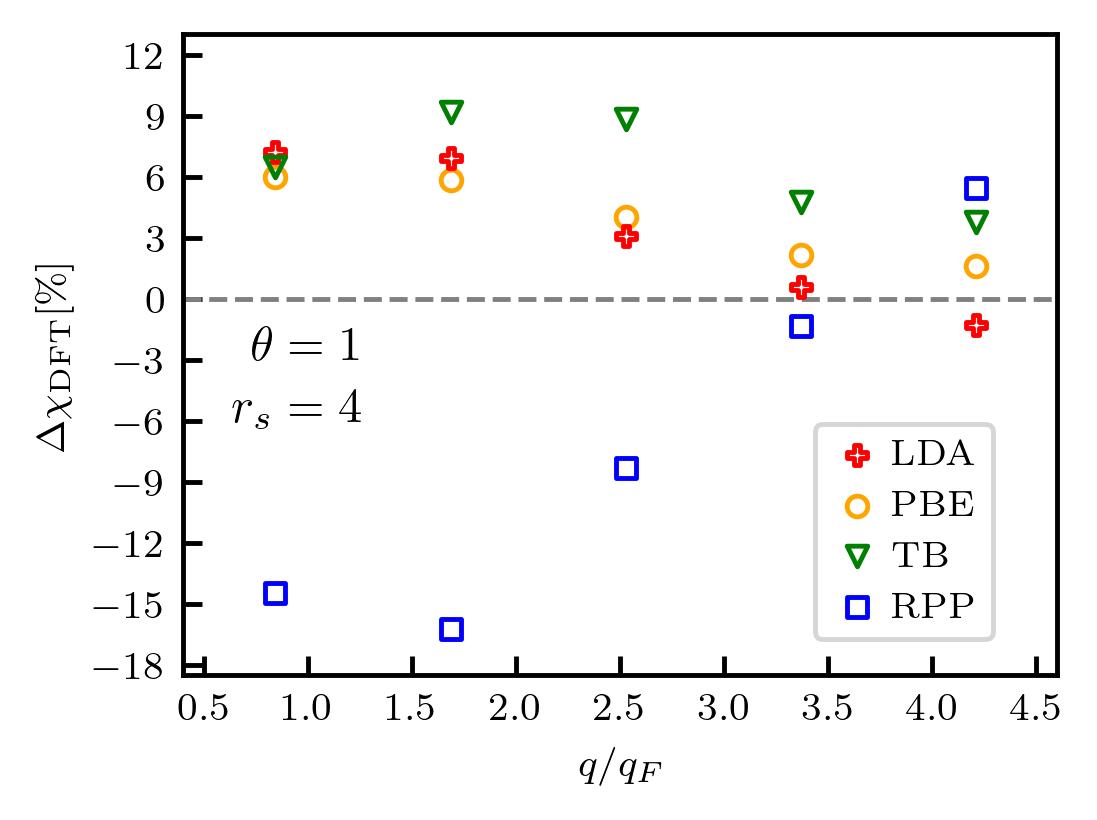}
\caption{\label{fig:chi_err_rs4} The deviation of the KS-DFT results from the exact PIMC data for the static density response function of warm dense hydrogen at $r_s=4$ and $\theta=1$. 
}
\end{figure}

We further quantify the difference between the KS-DFT data and the PIMC results using $\Delta \chi_{\rm DFT}$ defined by Eq.~(\ref{eq:dft_vs_pimc}).
The results for the relative difference $\Delta \chi_{\rm DFT}$ are presented in  Fig. \ref{fig:chi_err_rs4}.
From  Fig. \ref{fig:chi_err_rs4}, we observe that the LDA and PBE based data have similar quality with the largest deviation about $7\%$ at $q\simeq 0.845 q_F$.
The quality of the LDA and PBE based data improves with the increase in the wave number  with a relative disagreement of a few percent at $q\gtrsim 2.5q_F$ . Compared to the LDA and PBE, the results computed using the TB and RPP approximations are significantly worse and cannot be described as reliable at all considered wave numbers.

 \begin{figure}\centering
\includegraphics[width=0.45\textwidth]{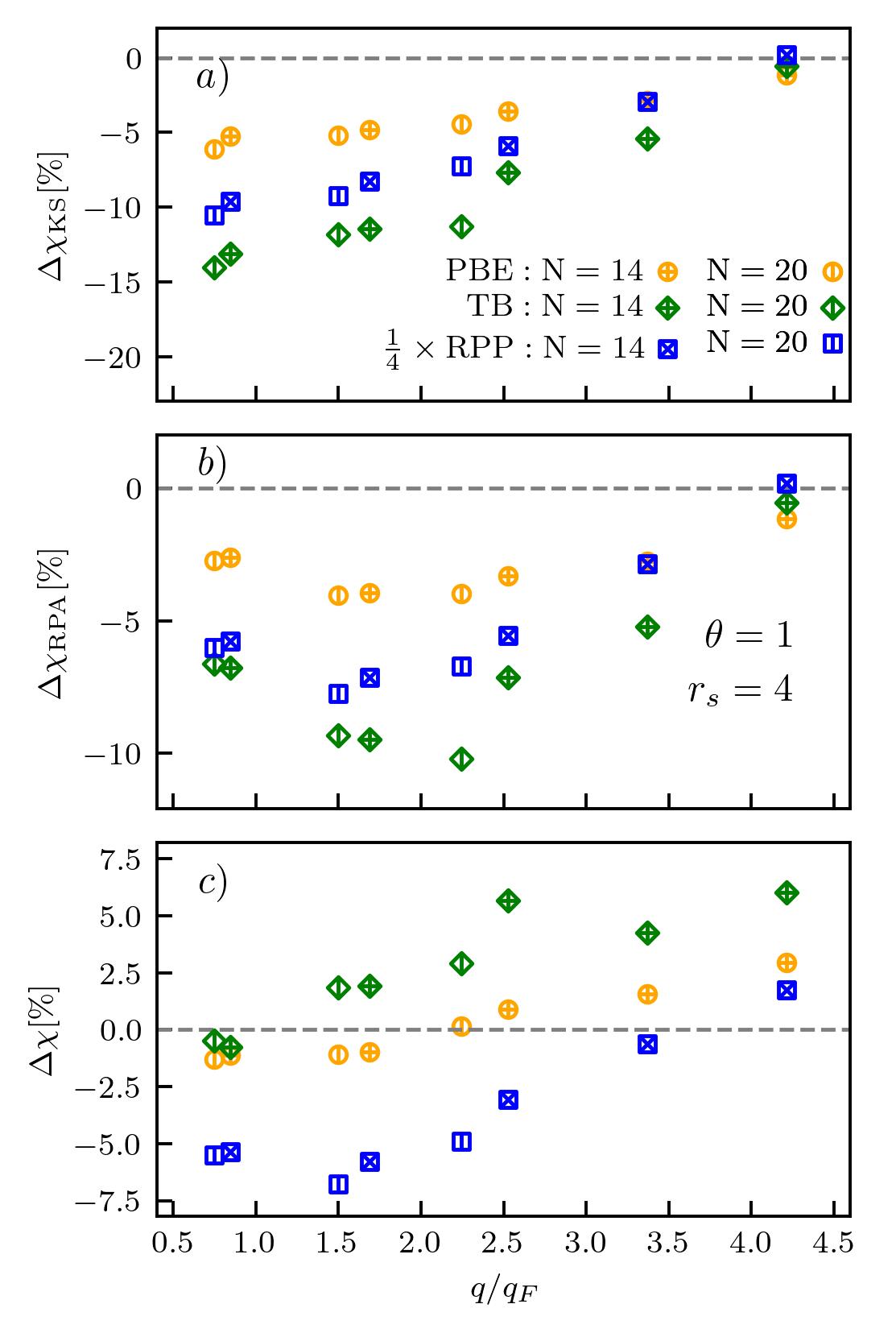}
\caption{\label{fig:ks_rs4} The relative deviation from the LDA results for a) the KS response function, b) the screened response on the level of RPA, and c) the  total density response function from simulations using GGA level PBE functional, meta-GGA level TB and RPP approximations for warm dense hydrogen at  $r_s=4$ and $\theta=1$. 
}
\end{figure} 

Let us now consider the effect of the approximation used for the XC potential on the  $\chi_{\rm KS}(q)$  and $\chi_{\rm RPA}(q)$ in more detail.
We present the relative deviation of the $\chi_{\rm KS}(q)$ and $\chi_{\rm RPA}(q)$ computed using the PBE, TB, and RPP compared to the LDA based data in Fig. \ref{fig:ks_rs4}a) and Fig. \ref{fig:ks_rs4}b), respectively. For the calculation of the relative deviations we used Eq.~(\ref{eq:diff_ks}) and Eq.~(\ref{eq:diff_rpa}). For completeness,  in Fig. \ref{fig:ks_rs4} we show the results computed for both 20 and 14 particles in the main cell. We note that in Fig. \ref{fig:ks_rs4}, the values of the RPP data points are reduced by a factor of 4 for a better illustration of the results.  
From  Fig. \ref{fig:ks_rs4}, one can see that the PBE and LDA based results for  $\chi_{\rm KS}(q)$  have a difference of about $5\%$ at $q<2.5q_F$. This difference monotonically reduces with the increase of the wave number at $q>2.5q_F$. Similar behavior is observed from Fig. \ref{fig:ks_rs4}b)  for $\chi_{\rm RPA}(q)$, but with the magnitude of the difference significantly smaller due to screening.  The TB and RPP based results for   $\chi_{\rm KS}(q)$  and $\chi_{\rm RPA}(q)$ are in agreement with the LDA based data only at large wave numbers $q>3q_F$ and drastically differ from the LDA based results at $q<3q_F$.

In  Fig. \ref{fig:ks_rs4}c), we present the relative deviation of the total $\chi(q)$ obtained using the PBE, TB, and RPP approximations compared to the LDA based results (computed using Eq.~(\ref{eq:diff_tot_chi})). The first interesting observation is that $\chi(q)$ computed using the TB is in good agreement with the LDA based result at $q<2.5q_F$ (with the difference $\Delta \chi (q)\lesssim 2.5~\%$ ). This is in contrast to the discussed relative differences for  $\Delta\chi_{\rm KS}(q)$ in Fig. \ref{fig:ks_rs4}a). 
Additionally, the $\Delta \chi (q)$ for the TB based results  increases with the increase in the wave number, while $\Delta\chi_{\rm KS}(q)$  decreases. 
This clearly demonstrates that $\chi_{\rm KS}(q)$---being an auxiliary  quantity---should not be used to gauge the quality of a particular XC functional upon comparing with actual properties of a physical system. 
The PBE based data for $\Delta \chi (q)$ show a good agreement with the LDA based results ($\Delta \chi (q)\lesssim 2.5\%$). Overall, the agreement between the PBE based $\chi(q)$ and the LDA based $\chi(q)$ is better than for $\chi_{\rm KS}(q)$  and $\chi_{\rm RPA}(q)$. From Fig. \ref{fig:ks_rs4}c), we see that the RPP approximation fails to describe the density response function of the warm dense hydrogen at $r_s=4$ and $\theta=1$ with $\Delta \chi (q)$ reaching about $28\%$. These conclusions are valid for both data sets with $N=14$ and $N=20$ particles in the main cell.

\begin{figure}\centering
\includegraphics[width=0.45\textwidth]{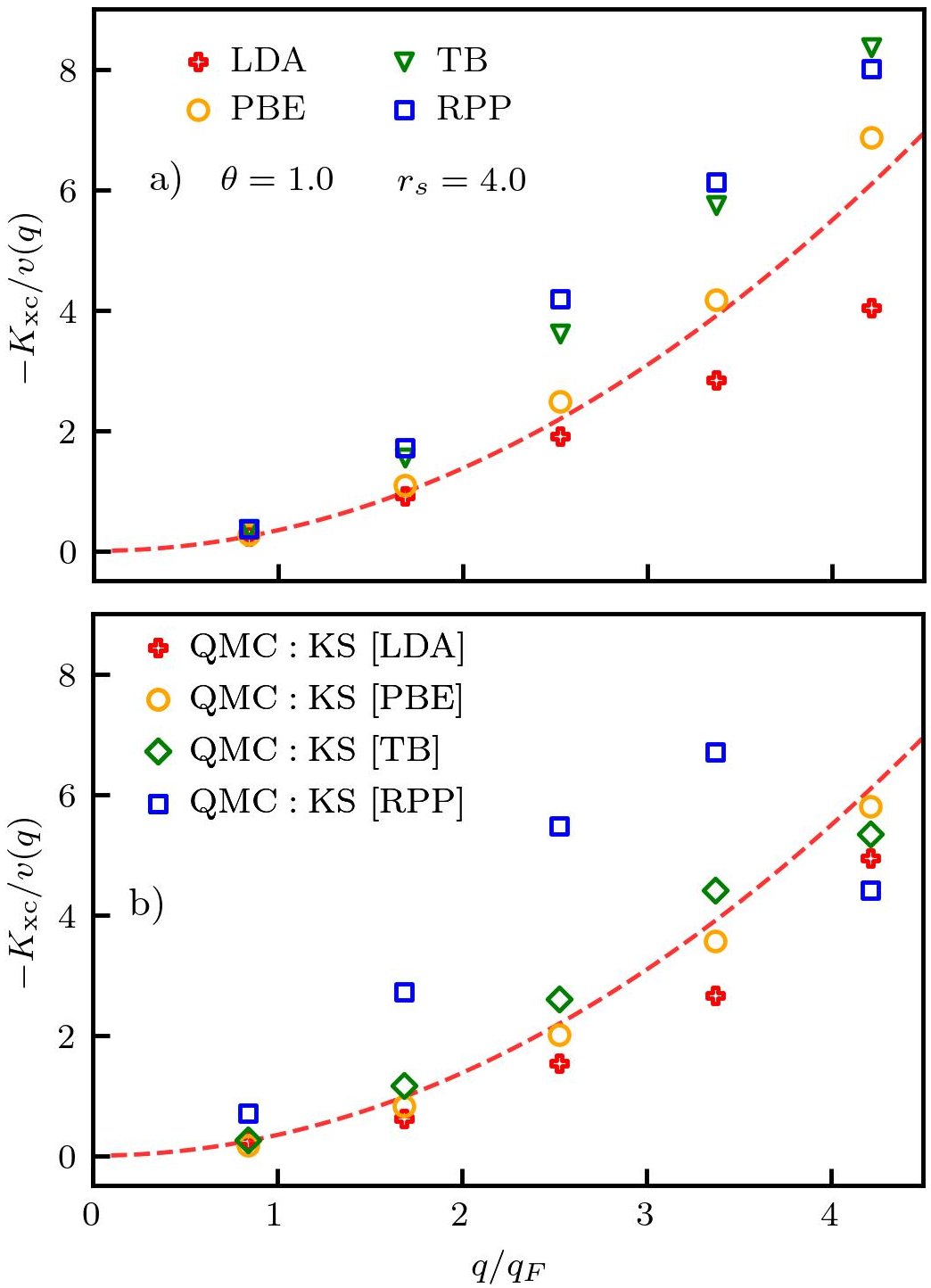}
\caption{\label{fig:LFCrs4} a) Fully consistent static XC-kernel from KS-DFT simulations with different XC functionals. b) Static XC-kernel extracted using the PIMC data and KS-response function from KS-DFT simulations. Demonstration of the equivalence of the KS response function from the harmonic perturbation technique and  LR-TDDFT calculations.  The results are for warm dense hydrogen at $r_s=4$ and $\theta=1$. 
}
\end{figure}

In Fig. \ref{fig:LFCrs4}, we show the results for the static XC kernel at $r_s=4$ and $\theta=1$. 
In Fig. \ref{fig:LFCrs4}a), we present the static XC kernel computed self-consistently from the KS-DFT. In Fig. \ref{fig:LFCrs4}b), we present the static XC kernel calculated using the PIMC data for $\chi(q)$.  The dashed red line in Fig. \ref{fig:LFCrs4} represents a quadratic dependence of the LFC $G(q)$ corresponding to the long-wavelength approximation. 

 \begin{figure}\centering
\includegraphics[width=0.45\textwidth]{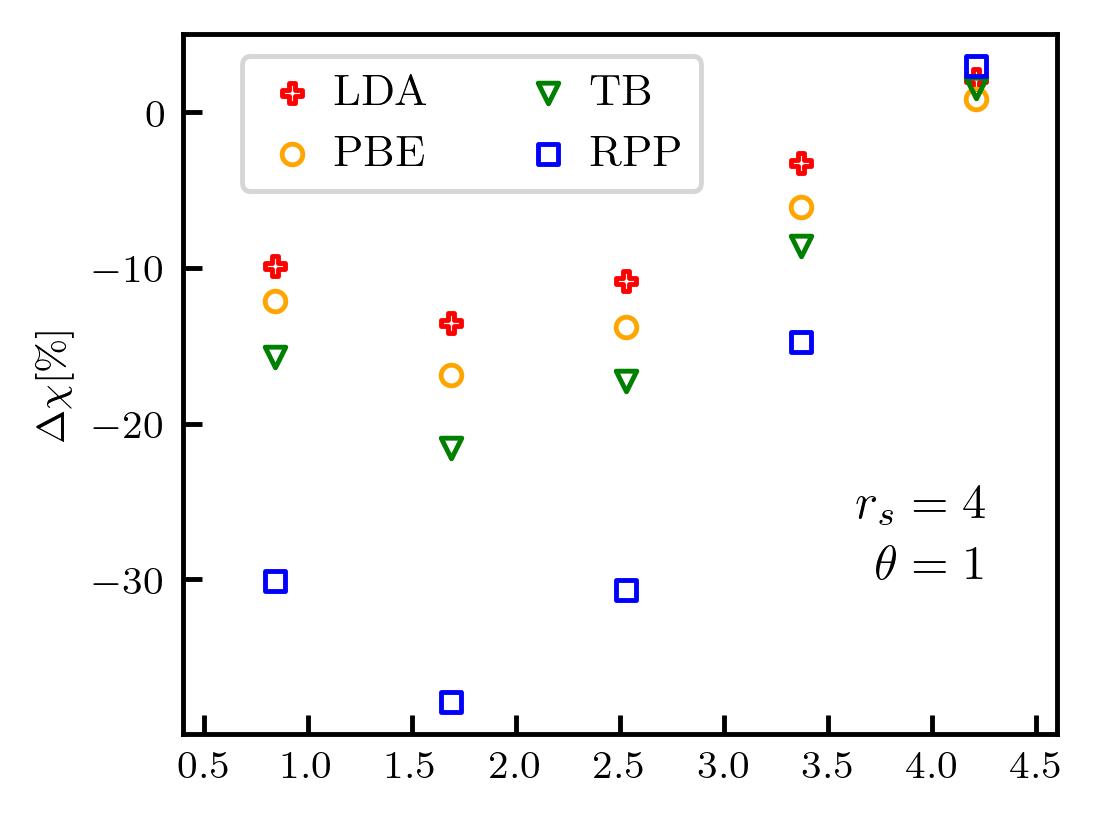}
\caption{\label{fig:NullXCrs4} Inaccuracy in the density response function due to an inconsistent combination of the KS response function and the XC kernel extracted from the PIMC data and an ideal reference function without XC effects. The results are for warm dense hydrogen at $r_s=4$ and $\theta=1$. 
}
\end{figure} 

From Fig. \ref{fig:LFCrs4}a), we see that the PBE based XC kernel is well described by the quadratic curve at $q<4q_F$. The same is the case for the LDA based XC kernel at $q\lesssim 2.5q_F$. The TB and RPP based XC kernels have significantly larger values than the PBE and LDA based data at $q>2q_F$. All considered XC kernels are in good agreement with each other at $q<q_F$.

In Fig. \ref{fig:LFCrs4}b),  we show the XC kernel computed using the PIMC data for $\chi(q)$. The PIMC XC kernel obtained using the PBE based $\chi_{\rm KS}(q)$ has values close to the  PBE based XC kernel from KS-DFT. This can be seen by comparing the PBE data positions relative to the dashed red line in Fig. \ref{fig:LFCrs4}a) and Fig. \ref{fig:LFCrs4}b).
This is also the case for the LDA based data. From Fig. \ref{fig:LFCrs4}b), we see that at $q>2q_F$, the PIMC XC kernel computed using the TB based $\chi_{\rm KS}(q)$ has significant differences compared to the data computed using the LDA and PBE based $\chi_{\rm KS}(q)$. The PIMC XC kernel computed using the TB based $\chi_{\rm KS}(q)$ is in good agreement with the PBE based data at $q<2q_F$. 

Let us now consider the performance of the XC kernel computed using the PIMC data for $\chi(q)$ and setting $\chi_{\rm KS}(q)=\chi_{\rm KS}^{\rm NXC}(q)$.
As for the case with $r_s=2$, we use the resulting static XC kernel in  Eq.~(\ref{eq:chi_adiabatic}) to compute $\chi(q)$. The obtained  $\chi(q)$ is compared  to the exact PIMC data for $\chi(q)$.
The corresponding results are shown in Fig. \ref{fig:NullXCrs4}. Comparing relative deviation (error) values presented in Fig. \ref{fig:NullXCrs4}  with Fig. \ref{fig:chi_err_rs4}, we clearly see that the use of the $\chi_{\rm KS}^{\rm NXC}(q)$ as a reference function for the extraction of the XC kernel from the PIMC data for  $\chi(q)$ leads to the significantly larger errors for the density response function even compared to the purely KS-DFT based XC kernel. It is clear that a strong localization of electrons around protons invalidates  the use of $\chi_{\rm KS}^{\rm NXC}(q)$ as a good reference function.

 \begin{figure}\centering
\includegraphics[width=0.45\textwidth]{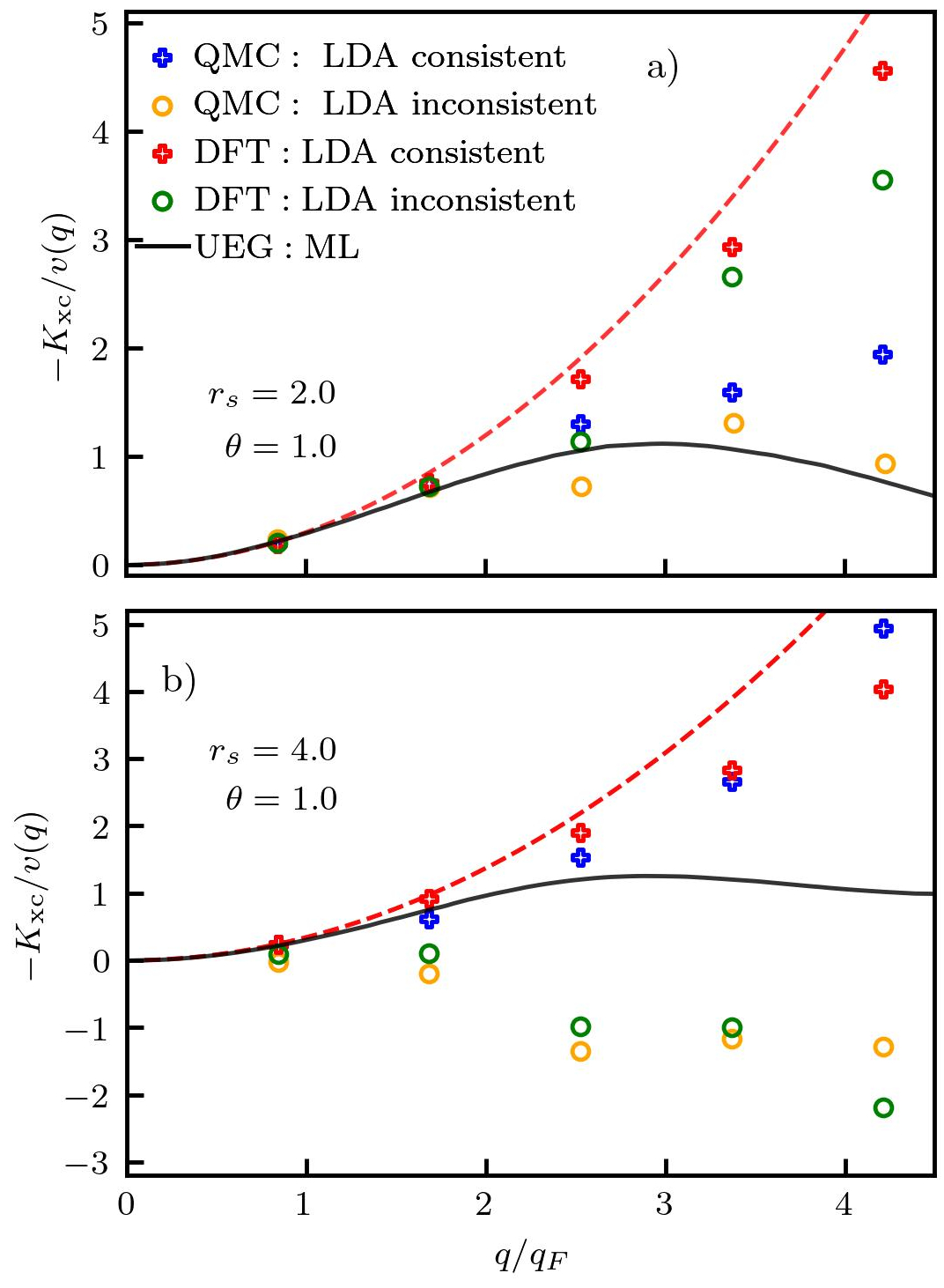}
\caption{\label{fig:PRL} The XC kernels of warm dense hydrogen computed using consistent and  inconsistent KS response functions at a) $r_s=2$ and b) $r_s=4$ with $\theta=1$.
We also show the exact UEG results based on the neural-net representation of Ref.~\cite{dornheim_ML} (solid black line).  
}
\end{figure} 

\subsection{The importance of a consistent KS response function for the static XC kernel}\label{s:prl}

The main goal of the paper is to show how to compute the  macroscopic KS response function that can be used together with the $\chi(q)$ for the calculation of the static  XC kernel and that can be used for the calculation of $\chi(q,\omega)$ consistently within the \textit{adiabatic approximation}, Eq.~(\ref{eq:chi_adiabatic}).
The standard way of introducing a macroscopic quantity in the LR-TDDFT at a wave number $\vec q=\vec k+\vec G$   is by setting $\vec G=\vec G^{\prime}$ \cite{book_Ullrich}  ($\vec G=0$, if $\vec q$ is in the first Brillouin zone). This is known to be valid for the dielectric function as we also confirm it here independently using the method of direct perturbation.
However, setting $\vec G=\vec G^{\prime}$ in  a non-interacting density response function $\chi^{~0}_{M}(q)=\chi^{~0}_{\scriptscriptstyle \vec G,\vec G}(\vec k,\omega)$ does not give a correct macroscopic KS response function. One of our key findings is that $\chi_{\rm KS}(q,\omega)\neq \chi^{~0}_{M}(q)$, where $\chi_{\rm KS}(q,\omega)$ is defined by Eq.~(\ref{eq:macro_ks}) or equivalently by Eq.~(\ref{eq:delta_KS}). We have been able to show that Eq.~(\ref{eq:macro_ks}) provides a correct macroscopic KS response function by using  Eq.~(\ref{eq:delta_KS}) within the direct perturbation approach. As far as we know, this is the first time that a proper macroscopic KS response function for disordered systems is defined. Now we can demonstrate the effect of an inconsistent $\chi^{~0}_{M}(q)$ on the extraction of the XC kernel from  $\chi(q)$. 

In Fig. \ref{fig:PRL} we compare the static XC kernels extracted from the PIMC and KS-DFT data for  $\chi(q)$ using $\chi_{\rm KS}(q)$ (a consistent macroscopic KS response function) and $\chi^{~0}_{M}(q)$  (an inconsistent  macroscopic KS response function). For the KS-DFT data in Fig. \ref{fig:PRL} we used the LDA functional. From Fig. \ref{fig:PRL}a), we see that at $r_s=2$ and $\theta=1$, 
the $\chi_{\rm KS}(q)$ and $\chi^{~0}_{M}(q)$ based data are  in agreement with each other at $q<2q_F$ and significantly differ from each other at $q>2q_F$.
At $r_s=4$ and $\theta=1$, we see from  Fig. \ref{fig:PRL}b) that   the $\chi_{\rm KS}(q)$ and $\chi^{~0}_{M}(q)$ based results for the XC kernel differ not only quantitatively, but also qualitatively. Indeed, the $\chi_{\rm KS}(q)$ based XC kernel is always positive, while the XC kernel computed using $\chi^{~0}_{M}(q)$  has negative values. 
This qualitative disagreement was interpreted in Ref. \cite{Bohme_PRL_2022} as a failure of the adiabatic LDA. Here we show that it is in fact not the failure of the adiabatic LDA, but caused by the use of the $\chi^{~0}_{M}(q)=\chi^{~0}_{\scriptscriptstyle \vec G,\vec G}(\vec k,\omega)$ as a macroscopic KS response function in Eq.~(\ref{eq:invert}).

\section{Conclusions and Outlook}\label{s:end}

Recently, with the focus on disordered systems, Moldabekov \textit{et al.} \cite{Moldabekov_dft_kernel} have introduced the direct perturbation based method for the computation of the static XC kernel for any available XC functional.  Here we show how this static XC kernel can be used for the LR-TDDFT calculations of the dynamic density response function $\chi(q,\omega)$ within the  \textit{adiabatic approximation} (represented by Eq.~(\ref{eq:chi_adiabatic}) and Fig. \ref{fig:scheme}). The main point is to use a proper and consistent macroscopic KS response function as introduced by Eq.~(\ref{eq:macro_ks}).  We reiterate that the presented adiabatic approximation for $\chi(q,\omega)$ has full consistency between the XC functional used for the computation of the XC kernel and for the KS response function. We have demonstrated how the inconsistent combination of the KS response function and XC kernel can lead to the quantitatively and even qualitatively wrong results. Additionally, we conclude from the performed analysis that the existence of a universal static XC kernel that works for different parameters is highly unlikely. Instead, one has to use a state and material specific static XC kernel, which can be computed using the direct perturbation approach. 

Furthermore, we have studied in detail the effect of the approximations made to the XC potential on the density response function and XC kernel of warm dense hydrogen.
We demonstrated: a) the application of  the presented  scheme for a static XC kernel that is self-consistent with the reference KS response function, b) the role of this self-consistence in the description of the static density response function, c)  the analysis of the static density response function using the exact PIMC data for warm dense hydrogen, and d) the role of the variation in XC functional on the macroscopic KS response function and XC kernel.

We found that at $\theta=1$,  the TR and RPP meta-GGA level approximations perform worse than the ground state LDA and PBE functionals.
Together with the prior finding of inefficiency of the meta-GGA level SCAN functional at WDM conditions \cite{Moldabekov_dft_kernel}, the problems of the TP and RPP for partially degenerate electrons indicate that 
the standard recipe for the construction of the XC functionals starting on the basis of the LDA does not lead to a better description of WDM. Indeed, even at the LDA level, an explicit inclusion of the temperature dependence into the XC functional leads to a worsening of the KS-DFT results quality at WDM conditions \cite{Moldabekov_dft_kernel}.
Therefore, it is clear that XC functional development for the WDM regime requires  new innovative approaches.
For example, it was recently shown that PBE0 type hybrid level functionals, mixing exact Hartree-Fock  exchange with the PBE exchange, can provide a good description of the UEG at WDM parameters if a mixing coefficient is chosen to reproduce the XC kernel of the UEG  \cite{hybrid_results, Moldabekov_non_empirical_hybrid}.
Now, using the consistent approach presented  in this work, this analysis of the PBE0 type hybrid level functionals can be extended to real materials like warm dense hydrogen.

An important possible application of the presented consistent adiabatic approximation for $\chi(q,\omega)$ is the  analysis and modelling of the XRTS measurements from WDM~\cite{Preston, GFGN2016:matter}. This is possible since the knowledge of $\chi(q,\omega)$  gives one straightforward access to the the dynamic structure factor $S_{ee}(\mathbf{q},\omega)$ of electrons via the fluctuation--dissipation theorem~\cite{quantum_theory}. Besides, the knowledge of $\chi(q,\omega)$ is equivalent to the knowledge of the dynamical dielectric function $\varepsilon(q,\omega)$, which in turn can be used for the calculation of transport properties like electrical conductivity \cite{Hamann_PRB_2020} and stopping power \cite{Moldabekov_PRE_2020}. 

Finally, we note that a consistent static XC kernel is needed for various other applications such as  the construction of effective potentials~\cite{Dornheim_JCP_2022, ceperley_potential,zhandos2,quantum_theory}, for quantum hydrodynamics~\cite{Murillo,zhandos_pop18,Moldabekov_SciPost_2022, Graziani_CPP_2022, zmcpp2017} and plasmonics \cite{Fabio_JCP},
and the for computation of the energy loss characteristics of high-energy density plasmas~\cite{Cortez_2018,  issanova_meister_2016, Ding_PRL_2018}. 

\section*{Acknowledgments}
This work was funded by the Center for Advanced Systems Understanding (CASUS) which is financed by Germany’s Federal Ministry of Education and Research (BMBF) and by the Saxon state government out of the State budget approved by the Saxon State Parliament. We gratefully acknowledge computation time at the Norddeutscher Verbund f\"ur Hoch- und H\"ochstleistungsrechnen (HLRN) under grant shp00026, and on the Bull Cluster at the Center for Information Services and High Performance Computing (ZIH) at Technische Universit\"at Dresden.




\bibliography{bibliography.bib}

\end{document}